\documentclass[a4paper,useAMS,usenatbib]{mn2e}
\usepackage{graphicx}
\usepackage{subfigure}

\newcommand{\my}{M$_{\odot}$\,yr$^{-1}$}
\newcommand{\ms}{M$_{\odot}$}

\long\def\symbolfootnote[#1]#2{\begingroup%
\def\thefootnote{\fnsymbol{footnote}}\footnote[#1]{#2}\endgroup}

\title[Wind-Capture Disks In Binary Systems]
{The Formation and Evolution of Wind-Capture Disks In Binary Systems}

\author[M. Huarte-Espinosa et al.]{M.~Huarte-Espinosa$^{1}$\thanks{E-mail: martinhe@pas.rochester.edu},
J.~Carroll-Nellenback$^{1}$, 
J.~Nordhaus$^{1,2,3}$
A.~Frank$^{1}$, 
E.~G.~Blackman$^{1}$ \\
$^1$Department of Physics and Astronomy, University of Rochester, 600 Wilson Boulevard,
Rochester, NY, 14627-0171 \\
$^2$NSF Fellow, Center for Computational Relativity and Gravitation, Rochester Institute of Technology, Rochester, NY 14623, U.S.A. \\
$^3$National Technical Institute for the Deaf, Rochester Institute of Technology, Rochester, NY 14623, U.S.A \\
}
\begin{document}

\date{Received \today}
\pagerange{\pageref{firstpage}--\pageref{lastpage}} \pubyear{2013}
\maketitle
\label{firstpage}

\begin{abstract}
We study the formation, evolution and physical properties of accretion
disks formed via wind capture in binary systems.  Using the AMR
code AstroBEAR, we have carried out high resolution 3D simulations
that follow a stellar mass secondary in the co-rotating frame as
it orbits a wind producing AGB primary.  We first derive a resolution
criteria, based on considerations of Bondi-Hoyle flows, that must
be met in order to properly resolve the formation of accretion disks
around the secondary.  We then compare simulations of binaries with
three different orbital radii ($R_o =\,$10, 15, 20\,AU). 
Disks are formed in all three cases, however the size of the disk
and, most importantly, its accretion rate decreases with orbital
radii.  In addition, 
the shape of the orbital motions of
material within the disk becomes increasingly elliptical with
increasing binary separation.  The flow is mildly unsteady with
``fluttering'' around the bow shock observed. The disks are generally
well aligned with the orbital plane after a few binary orbits. We
do not observe the presence of any large scale, violent instabilities
(such as the flip-flop mode). 
For the first time, moreover, it is observed 
that the wind component that
is accreted towards the secondary has a vortex tube-like
structure, rather than a column-like one as it was previously thought.
In the context of AGB binary systems
that might be precursors to Pre-Planetary and Planetary Nebula,
we find that the wind accretion rates at the chosen orbital separations are
generally too small to produce the most powerful outflows observed
in these systems if the companions are main sequence stars but
marginally capable if the companions are white dwarfs.  It is likely
that many of the more powerful PPN and PN involve closer binaries
than the ones considered here.  
The results also demonstrate principles of
broad relevance to all wind-capture binary systems.
\end{abstract}


\section{Introduction}

The formation of accretion disks in binary stars can occur in 
several 
ways.  If the orbital separation of the stars is small enough then
the donor star can overflow its Roche lobe and transfer material
to the accreting star where conservation of angular momentum will
force the development of an accretion disk.  This process has been
well studied both analytically and via simulations (e.g.
Eggleton 1983; D'Souza et al. 2006).  The second mechanism, wind
capture via Bondi-Hoyle-Lyttleton (BHL) flows, is less well studied
and fundamental questions as to its efficacy remain. 
In particular it is not clear how large an orbital separation is
possible that still allows for wind capture accretion disks to form.
This question applies for a range of  binary accretion systems but
our particular focus is on  the late stages of evolution for low-
and intermediate-mass stars.

Over the last decade the traditional paradigm for planetary nebula
formation, in which successive winds from a single star create the
nebula, has been challenged on a number of theoretical and
observational
fronts (Bujarrabal et al. 2001; Soker \& Rappaport 2000, 2001;
Balick \& Frank 2002;  Nordhaus et al. 2007; De Marco 2009; Witt
et al. 2009; Huarte-Espinosa et al. 2012; 
Soker \& Kashi 2012).  If indeed binaries play a
central role in the formation of pre-planetary and planetary nebula
(PPN/PN) then we must revise our understanding of the evolutionary
paths for sun-like stars.  The basic morphological influence of a
secondary on the wind poses one set of questions (e.g. Edgar et al.
2008; Huggins et al.~2009; Kim \& Taam 2012ab), but  the formation
of accretion disks around the secondary is of more direct importance
for constraining which subset of specific binary accretion scenarios
(Reyes-Ruiz \& Lopez 1999; Soker \& Rappaport 2000,2001; Blackman
et al. 2001)
can lead to the powerful collimated outflows observed in PN and
PPN (Bujarrabal et al. 2001).  Since accretion disks  plausibly
drive collimated outflows via magneto-rotational processes 
(\citealp{blandford82}; \citealp{ouyed97};
\citealp{blackman01}; \citealp{mohamed07})
determining the conditions for disk formation remains a
basic issue for studies of the low/intermediate mass star late
evolutionary stages.  In particular, understanding the range of
orbital separations in which disks can form with sufficient accretion
rates  is important for  population synthesis models that can
determine which systems will become PN and/or PPN.

The subject is analytically and computationally subtle and there have
been only a handful of  studies  addressing the aforementioned
questions. Thuens \& Jorisen (1993)
carried forward early work
on the subject using a Smooth Particle Hydrodynamics (SPH) code. 
Mastrodemos \& Morris (1998,99;
which we will refer to as MM98 and MM99, respectively) 
also used SPH simulations to demonstrate that disks
can form in AGB specific binary systems. Even though the resolution
was low by contemporary standards they were able to set some limits
on where disks could form and gain insight into a limited set of
disk characteristics.  In Soker \& Rappaport~(2000) analytic estimates
were given for when disks would form within AGB binary systems. 
General studies of wind accretion 
in binaries using a grid based code were carried out by Nagae et
al.~(2004) and Jahanara et al.~(2005), 
who explored mass accretion rates and
angular momentum losses. Recently, the issue of disks in AGB binary
systems has been taken on numerically by de~Val-Boro et al.~(2009) who carried
forward high resolution simulations using the AMR code FLASH.
Such work was carried out in 2-D and showed that, in this
restricted geometry, disks might form for binary separations out to 40\,AU. Mohamed \& 
Podsiadlowski~(2007) 
also studied the problem, but using the code GADGET, with a focus
on the limits of Roche lobe overflow. Disk
formation and structure have also been explored via semi-analytic models
by Perets \& Kenyon~(2012) who found disks forming out to 100\,AU.

Here we carry out the first fully 3-D high resolution simulation
study BHL flows and disk formation in AGB binary systems since MM98.  
Our goals
are to explore the mechanisms of disk formation, to study the flow
features associated with fully formed disks in binaries, and to
obtain a first cut at the properties of disks formed in these
systems.  In addressing all of these issues, we are specifically
interested in systems relevant to PPN and PN, but our results will
be broadly applicable to wind capture binaries.

The structure of our paper is as follows.  In section \ref{bondi},
we briefly discuss BHL flows in binaries in order to derive a
resolution condition that should be met to accurately track
the formation of disks in wind capture systems.  In section \ref{model}
we discuss the methods and models used in our study.  Section
\ref{results} provides 
a description and analyses of the results. We then
discuss accretion and its implications for PPN and PN in section~\ref{discu}.
Finally, in section
\ref{conclu} we provide our conclusions and their relevance to
PPN/PN, BHL studies and binary systems as a whole.

\section{Numerical Resolution Criterion for Binary BHL Flows} \label{bondi}

We begin with an evaluation of 
BHL accretion in the context
of binary stars.  A excellent review of BHL physics in general can
be found in \citet{edgar04}.  Our goal in this section is to
characterize a resolution criterion for tracking the formation and
evolution of disks in BHL binary systems.

The general BHL problem occurs when a compact object of mass, $m$,
moves at a constant supersonic relative velocity, $v_r$, through an infinite
gas cloud which is homogeneous at infinity.  The gravitational field
of $m$ focuses the cloud material located within the BHL radius

\begin{equation}
  r_b = 2Gm/v_r^2.
  \label{rb}
\end{equation}

\noindent The focused material does not accrete directly onto the
object but forms a downstream wake.  A conical shock is formed along
the so called ``accretion line'' 
(or accretion column) 
connecting the wake and the object.
This flow becomes divided into (i) material accreted onto the object
and (ii) material that flows away and escapes from the object.
These flow components are separated by a stagnation point (Bondi
\& Hoyle, 1944).

\begin{figure}
\centering
  \includegraphics[width=\columnwidth]{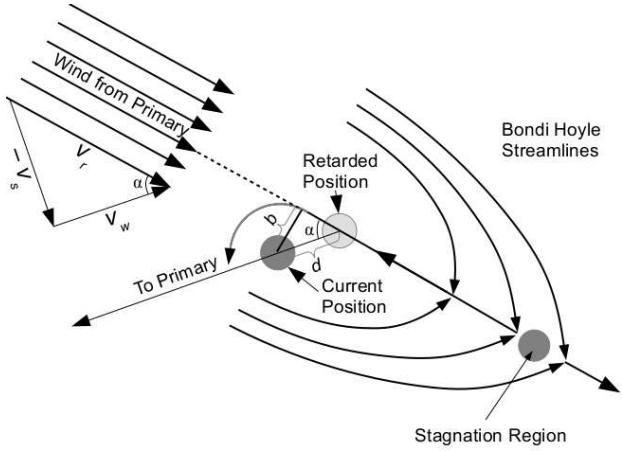}
  \caption{Schematic showing the flow structure of the binary wind capture
  process.
		 The figure is in the inertial frame of the secondary and the
		 direction of the motion of the primary is to the top left.
       The unlabeled curved arrow indicates the accretion column bending around 
		 the secondary and forming a disk.
  }
  \label{impactParameter}
\end{figure}

Here, we study a system in which the cloud material is that of an
AGB star's wind moving radially outward with a terminal speed, $v_w$.
The accretor is an orbiting companion of mass, $m_s$, with an orbital
separation of $R\ge\,$10\,AU and orbital velocity, $v_s$. The relative
wind speed seen by the companion is therefore $v_r=\sqrt{v_w^2+v_s^2}$.
One essential difference between the classic BHL analysis and the
binary case is the accretion flow in the wake.  
As a consequence of the systems orbital motion, gas at the stagnation point
does not flow directly toward the secondary but is aimed slightly off axis.  This
occurs because the instantaneous position of the companion relative
to the BHL accretion flow is constantly changing. Hence material in the wake flows
towards a point between the companion's earlier and current positions;
i.e. the focused material is accelerated towards a retarded position.
Figure~\ref{impactParameter} shows the considered flow structure.
It is, in fact, this off-center streaming of the accretion flow
(and its associated conservation of angular momentum) that allows
an accretion disk to form. Resolving this process 
   numerically
is therefore
critical to capturing the formation of an accretion disk.  We note
that from the perspective of a reference frame co-rotating with the
companion, the shifting
of the accretion stream occurs due to non-inertial forces.  

\begin{figure}
\centering
  \includegraphics[width=.7\columnwidth, angle=270]{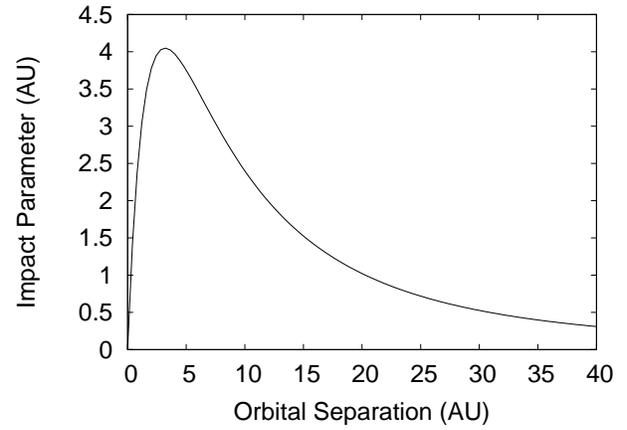}
  \caption{Plot showing 
   the accretion line
	impact parameter as a function of separation
  for a 1.5\,\ms\ primary, a 1\,\ms\ secondary 
      and a wind with terminal speed of 10\,km\,s$^{-1}$.}
  \label{impact_curve}
\end{figure}

The time scale, $t_c$, associated with the wind capture process scales
approximately with the time scale for material to flow through the
BHL radius, that is $t_c = \frac{r_b}{v_r} = \frac{2Gm_s}{v_r^3}$,
where (again) $v_r$ is the velocity of the gas relative to the companion.
In an inertial frame moving with the secondary's velocity at $t_0$,
the flow resembles a classical BHL pattern, although the secondary
is accelerated from rest toward the primary with $a_c=\frac{v_s^2}{r_s}$.
The distance, $d$, travelled by the secondary during the wind capture
time scale $t_c$ is 

\begin{equation}
   d = \frac{1}{2} a_c t_c^2,
\label{d_sec}
\end{equation}
\noindent and the distance perpendicular to the accretion 
   column, 
is 

\begin{equation}
   b = d \sin{\alpha} = d \frac{v_s}{v_r}.
\label{par2.5}
\end{equation}
\noindent We call $b$ the 
   accretion line
impact parameter which can be expanded as
 \begin{equation}
b = \frac{1}{2} \frac{v_s^2}{r_s} \left( \frac{2Gm_s}{v_r^3} \right)^2
\frac{v_s}{v_r} = \frac{2 G^2 m_s^2 v_s^3}{r_s v_r^7}.
\label{par3}
\end{equation}
\noindent Solving for $v_s$ and $r_s$ in terms of the mass of the
primary, $m_p$, the total mass, $M=m_s+m_p$, and the total separation,
$R$, we get

\begin{equation}
b= 2 \left ( \frac{m_s^2M^3}{m_p^5} \right )\frac{\left( \frac{1}{R} 
\right)^{5/2}}{\left( \frac{1}{R} + \frac{1}{r_w} \right)^{7/2}}
\label{b_is}
\end{equation}
\noindent where $r_w = \frac{Gm_p^2}{Mv_w^2}$ which for our choice
of wind velocity and masses gives the profile in
Figure~\ref{impact_curve}.

We note two important points. First, the expression for (\ref{b_is})
is valid only for timescales short enough for the frame to be
considered inertial. Thus our expression for the accretion line
impact parameter only holds for timescales short compared to the
orbital time ($t << P$, where $P$ is the orbital period).  This is
also the same limit as $d << R$. Taken together these limits also
imply our expression is only valid when the orbital separation $r_s$
is greater than the Bondi-Hoyle radius $r_b$.

In addition we note that our model implementation (section~\ref{model})
assumes that the wind experiences a balance of radiation pressure and
gravitational forces from the primary. This is equivalent to
assuming the secondary orbits beyond the wind's acceleration zone.
Operationally this means a differential acceleration between the
wind and the secondary occurs since the secondary does feel the
full gravity from the primary. It is this differential acceleration
which creates the accretion line impact parameter.  We note however
that even in the wind acceleration zone the radiation pressure force
on the gas (but not on the secondary) would produce such a differential
gravitational acceleration.

Finally we note that characterizing the appropriate ``disk formation''
radius in the context of Bondi-Hoyle flow in binaries is nontrivial.
Different approaches have been used by different authors
(\citealp{Wang81}).  For example, when considering flows with density
gradients of scale $H$, \citet{SL84} used earlier results by
\citet{DoddMcCrea52}, and derived an expression for disk formation
radius to be $r_k=GM/v^2 (r_b/2H)^2$.  In another work,
\citet{soker} compared the angular momentum of accreted
material through the Bondi-Hoyle radius $j_a$ to that of material
in Keplerian orbit around the accretor $j_2$ such that $j_2 < j_a$
for a disk to form.  This expression can also be used to derive an
expression for a disk formation radius $r_a$.  We find that our
expression for the accretion line impact parameter $b$ will, in the
region of its validity, be larger than both these radii.  This
indicates that $b$, again in its domain of validity, defines the
appropriate disk formation radius.

One of the primary 
   and novel
results of the simulations that we present here 
is showing that
accurately resolving this physical scale
in the vicinity of the secondary is essential to capturing the
dynamics of the flow.  Since the material initially goes into orbit
around the secondary with a radius of $b$, a failure to resolve the
scale can lead to the lack of an accretion disk forming or the
formation of disk with unphysical dynamics such as extreme warps
or tilts.

\section{Computational Methodology}%%%%%%%%%%%%%%%%%%%%%%%%%%%%%%%%%%%%%%%%
\label{model}

To model the formation and evolution of disks in binary system, we
use the adaptive-mesh-refinement (AMR), magnetohydrodynamical code
\textit{AstroBEAR}2.0\footnote{
https://clover.pas.rochester.edu/trac/astrobear/wiki}.  AstroBEAR
is a highly-parallel, second-order accurate, shock-capturing,
multi-physics Eulerian code \citep{astrobear,bear2}.
While AstroBEAR2.0 has options for MHD, self-gravity and heat
conduction, we neglect these processes in our current study.

We use \textit{BlueHive}\footnote{
http://www.circ.rochester.edu/wiki/index.php/BlueHive\_Cluster} 
--an IBM massively parallel processing
supercomputer of the Center for Integrated Research Computing of
the University of Rochester-- and \textit{Ranger}\footnote{
https://www.xsede.org/web/guest/tacc-ranger} --a Sun Constellation
Linux Cluster which is part of the TeraGrid project-- to run
simulations for an average running time of about 0.5\,days$/$orbit
using 64-1024 processors.

We perform calculations in three-dimensions and solve the equations of
hydrodynamics in a co-rotating frame, which in conservative form are:
\begin{equation}
\begin{array}{l l}
   \frac{\partial\rho}{\partial t} + \nabla \cdot(\rho\mbox{\bf v}) &= 0 \label{contin} \\
   \frac{\partial\left({\rho\bf v}\right)}{\partial t} +
   \nabla \cdot\left(\rho{\bf v v}\right) &= 
		-\nabla p - \rho\nabla\Phi -2\rho \Omega \times {\bf v}
		\label{contin2} \\
		~ & \quad - \rho \Omega \times \left ( \Omega \times {\bf r} \right )
\end{array}
\end{equation}
\noindent where $\rho$, $p$ and {\bf v} are the gas density,
pressure and flow velocity, respectively.  
  The last two terms in the right hand side of equation~(\ref{contin2}) 
  account for the centrifugal and the Coriolis forces that act on the gas.
We employ an isothermal ideal gas equation of state with an adiabatic
index of $\gamma=1$.

We follow the secondary star as it orbits about the system's center
of mass with a circular trajectory due to the gravitational field of the
binary system.
%The gravitational interaction of the two stars has been
%followed fully. They orbit each other in a circular trajectory about their
%center of mass. 
However, the gravitational effects that the
stars have on the gas is modelled to be caused by
the secondary star only. Thus we assume that there is a balance between the gravitational potential of
the primary star and the wind radiation pressure (Section~3.1.3).
The gravitational potential of the secondary star which
affects the gas is computed using
the functional form $ \Phi = G m_s / r $ for $r \ge 4 \sqrt{dx^2+dy^2+dz^2}$,
and we use a spline softening function at smaller radii.

\subsection{Initial Conditions and Set-up} %%%%%%%%%%%%%%%%%%%%%%%%%%%%%%%%%%%%%
\label{ini}

\begin{figure}
\centering
  \includegraphics[width=.7\columnwidth]{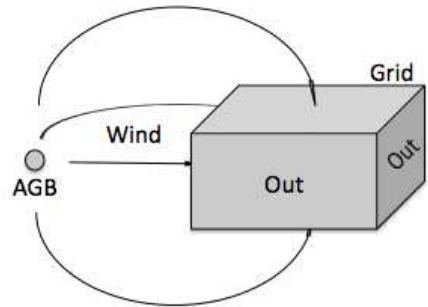} ~~
  \includegraphics[width=.7\columnwidth]{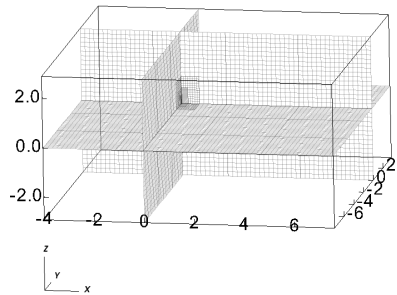}
  \caption{Top: Diagram showing the wind model and the grid.
  Bottom: Grid, its coordinates, in AU, and the central mesh. The secondary
  is located at the origin.}
  \label{diag}
\end{figure}

Our computational domain consists of a rectangular volume of
10x10x5\,AU$^3$.  We employ a fixed grid of 64x64x32~base cells
plus three, or four (see Table~1), AMR levels 
of factor two refinement. Nested refined cell
blocks are centered at the secondary's position such that the finest
level has 32-64 cells (on the orbital plane) enabling us to
resolve the central part of the disks and $b$ (section~2).
We note that in our implementation the primary star is located
outside the computational domain (see Figure~\ref{diag},
top).

We employ either wind boundary conditions (section~\ref{wind}) in
the $-x, +y$ and $\pm z$ domain faces, or outflow only conditions
in all other faces (Figure~\ref{diag}). %The positions of the primary,
%the center of mass and the secondary are ($-r_p$,0,0), (0,0,0) and
%($r_s$,0,0), respectively.  
We include Coriolis and centrifugal
terms and use a reference frame which co-rotates with the secondary.

\subsubsection{Primary} We use a simple AGB star model for the
primary consisting of a mass of 1.5\,M$_{\odot}$ and a wind with
an isotropic velocity of $v_w=\,$10\,km\,s$^{-1}$ and mass-loss
rate of $\dot{m}_p =\,$10$^{-5}$\,\my.  These values are typical
for AGB stars (e.g. Hrivnak et al., 1989; Bujarrabal et al., 2001).

We note a few aspects of this model. First, winds from AGB stars are
driven by radiation pressure acting on dust grains. Such interactions
are complicated and many details depend subtle non-linear effects \citep[][and references
therein]{sandin03}.  
Additionally, observations suggest that AGB-star winds possess
acceleration zones in which the radial wind velocity climbs slowly to its final value (e.g. see
\citealp{sandin}).  Such acceleration regions may play a role in
the formation of disks in binaries with separations $\la$\,10\,AU.
We expect this effect to be mild for the separations explored here
(10-20\,AU), thus no acceleration region has been considered 
and we leave more realistic AGB star wind models for future work.  
Additionally, we note that the binary separations that we 
explore here are outside
the wind-Roche-lobe-overflow (WROLF) limit \citep{sandin,mohamed11}.

\subsubsection{Companion} \label{sink} 
We model the secondary star using a sink particle, based on the
implementation of \citet{2004ASPC..323..401K}, which accretes gas
within a radius of 4 grid cells. The mass, momentum and energy of
the gas within this region are then acquired by the 
   secondary 
in a conservative fashion.  We set the 
   secondary 
with an initial mass of
1\,M$_{\odot}$ in a circular orbit about the primary.  While important
at these radii, we do not adjust the orbit for the effect of mass
loss or tidal interactions as they occur on longer evolutionary
timescales (Nordhaus et al.~2010; Nordhaus \& Spiegel 2012;
Spiegel \& Madhusudhan 2012).

\subsubsection{Wind solution and injection}
\label{wind}

We set the initial conditions throughout the grid with
a constant temperature of 1000\,K and a density given by

\begin{equation}
\rho_w = \frac{ \dot{m}_p }{ 4 \pi ({\bf x}_p-{\bf x})^2 v_w },
\label{densw}
\end{equation}
\noindent where $\dot{m}_p=\,$5$\times\,$10$^{-5}$\,M$_{\odot}$\,yr$^{-1}$,
$v_w=\,$10\,km\,s$^{-1}$, ${\bf x}_p$ and ${\bf x}$ are the positions
of primary's orbit and that of an arbitrary grid cell relative to
the center of mass, %(0,0,0), 
respectively.  We calculate the velocity
field of the solution by solving for the characteristics that leave
the surface of the primary at a \textit{retarded} time, $t_r=t-|{\bf
x}|/v_w$, with a velocity vector pointing towards ${\bf x}$. Wind
speeds are chosen so that $v_w > |{\bf v}_s|$, where ${\bf v}_s$
is the secondary's orbital velocity. We assume that the distance
from the primary's surface to ${\bf x}$ is larger than $|{\bf x}_p|$,
which constrains the separation between the secondary and the
grid's boundaries.
   The wind that arrives at point ${\bf x}$ at the current time $t$
	left the primary's surface at a retarded time $t_r$. The vector 
	connecting the primary's retarded position and our current position 
	${\bf x}$ is given by 

   \begin{equation}
      {\bf d} = {\bf x}-{\bf x}_p(t_r).
   	\label{3.1.3.d}
   \end{equation}
   \noindent At any point on the primary, the wind velocity field, 
	${\bf u}={\bf u}({\bf x},t)$, is given by 

   \begin{equation}
   {\bf u}=v_w {\bf \hat{n}}+{\bf v}_p(t_r).
   \label{wind2}
   \end{equation}
   \noindent The vectors ${\bf u}$ and ${\bf d}$ should be parallel at
	$t_r$ --when ${\bf u}$ left the primary's surface, i.e.
	
   \begin{equation}
   (v_w {\bf \hat{n}} + {\bf v}_p(t_r)) \times {\bf d} =0.
   	\label{3.1.3.cross}
   \end{equation}
   \noindent We can then solve for ${\bf \hat{n}}$ and the 
	wind vector at the retarded time. A better approximation of the retarded 
	time is then found using ${\bf u}$:
   \begin{equation}
   \tau_r=t-|{\bf x}|/|{\bf u}|.
   \label{wind3}
   \end{equation}

We iterate the above computations until we find a convergent wind
solution for the gird cell located at ${\bf x}$.  
Finally, we transform the solution to a
reference frame which co-rotates with the secondary.

	Once the initial conditions are set,
we continually give the above wind solution to the grid cells in the
$-x, +y$ and $\pm z$ domain faces 
(Figure~\ref{diag}, top).
For each iteration however we
use $\tau_r, \,{\bf x}_p(\tau_r)$ and ${\bf u}$ as initial conditions
in a recursive Runge-Kutta method to calculate the deflection of
${\bf u}$ --as it travels from the primary's surface [$\sim\,{x}_p(\tau_r)]$
to ${\bf x}$-- caused by the secondary's gravity field. We consider
the orbital motion of the secondary in these computations. This
step yields yet better estimates of both ${\bf \hat{n}}$ and $\tau_r$ than
the calculations in equation~(\ref{wind2}) alone.  We therefore use the
new values of ${\bf \hat{n}}$ and $\tau_r$ for the next iteration of
the wind solution.

Finally, to match the initial conditions and the injected wind
solution we allow the gravitational effect that the secondary has on the
gas to increase linearly during one wind crossing time, 1.6\,$r_b/v_w$,
from zero to $Gm_s/r$. The field remains constant thereafter.  This
has no effect on the binary orbital motion.

\section{Results}
\label{results}

We carry out three simulations corresponding to stellar separation
of 10, 15 and 20\,AU. Table~1 summarizes the models and their
relevant parameters.

%%%%%%%%%%%%%%%%%%%%%%%%%%%%%%%%%%%%%%%%%%%%%%%%%%%%%
\setcounter{table}{0}
\begin{table}
\centering
    \begin{minipage}{108mm}
\caption{Simulations and parameters. \label{t1}}
   \begin{tabular}{@{}lccccc@{}}
\hline           
Name   &Separation [AU]   &Resolution   
&$r_b$\footnote{BHL radius, equation~(\ref{rb}).} [AU] 
&$b$\footnote{
   Accretion line
impact parameter, equation~(\ref{b_is}).} [AU] \\
\hline           
M1     &10           	 &64x64x32+3amr    &4.9    &2.4 \\
M2     &15           	 &64x64x32+4amr		&5.8   &1.5\\
M3     &20           	 &64x64x32+4amr		&6.3   &1.0\\
   \hline
   \end{tabular}
   \end{minipage}
\label{table1}
\end{table}
%%%%%%%%%%%%%%%%%%%%%%%%%%%%%%%%%%%%%%%%%%%%%%%%%%%%

\begin{figure*}
\centering
$\log{(n)}$ [part\,cm$^{-3}$] 
$\qquad \qquad \qquad \qquad \qquad$ Flow speed [Mach] \\
\includegraphics[width=.55\textwidth,bb=030 577 415 610,clip=]{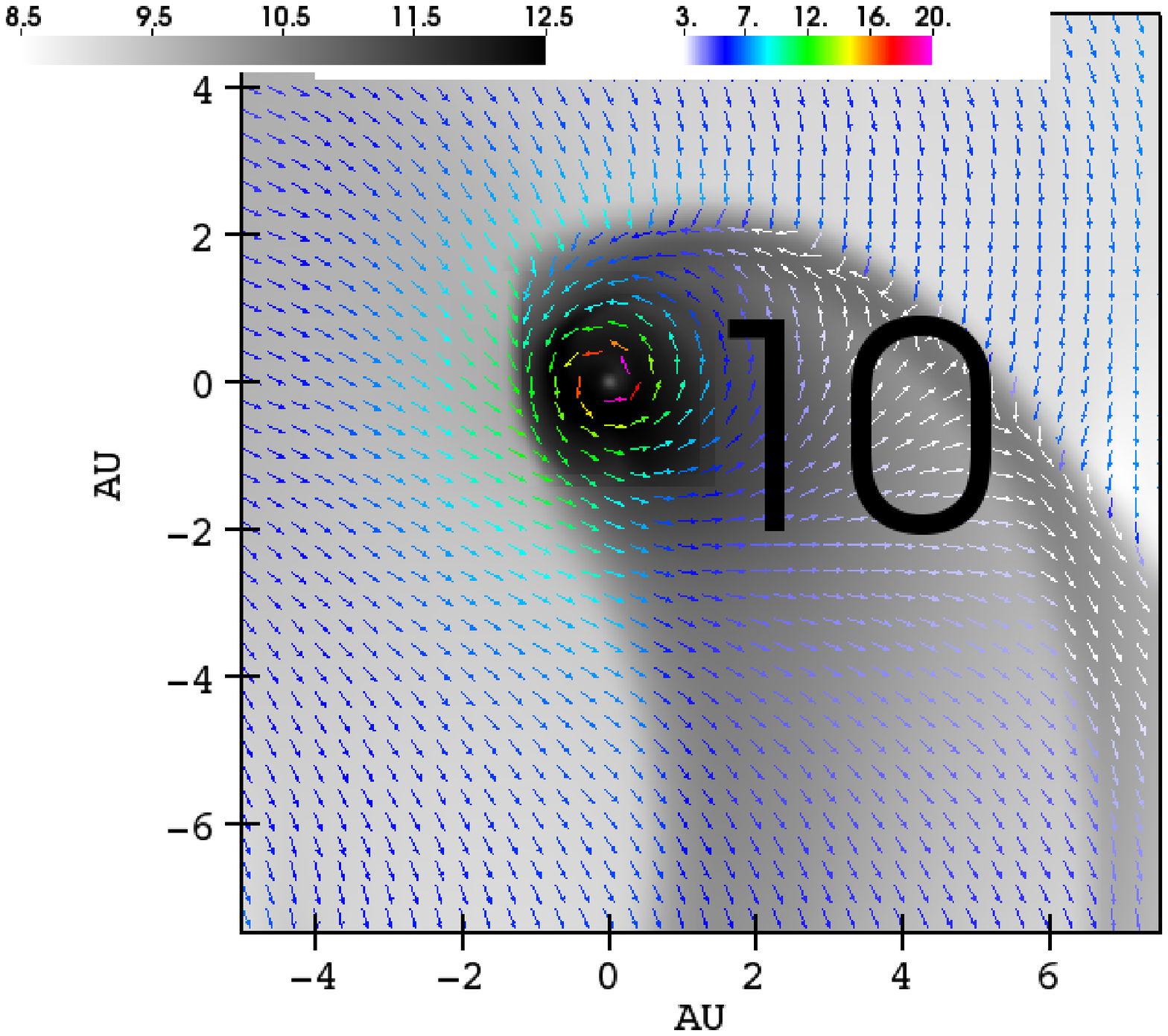}\\
~~~~~~~~~~~~~~~~~~~Face-on: \\
\vskip.10cm
     ~\includegraphics[width=.355\textwidth,bb= 40 175  580 660,clip=]{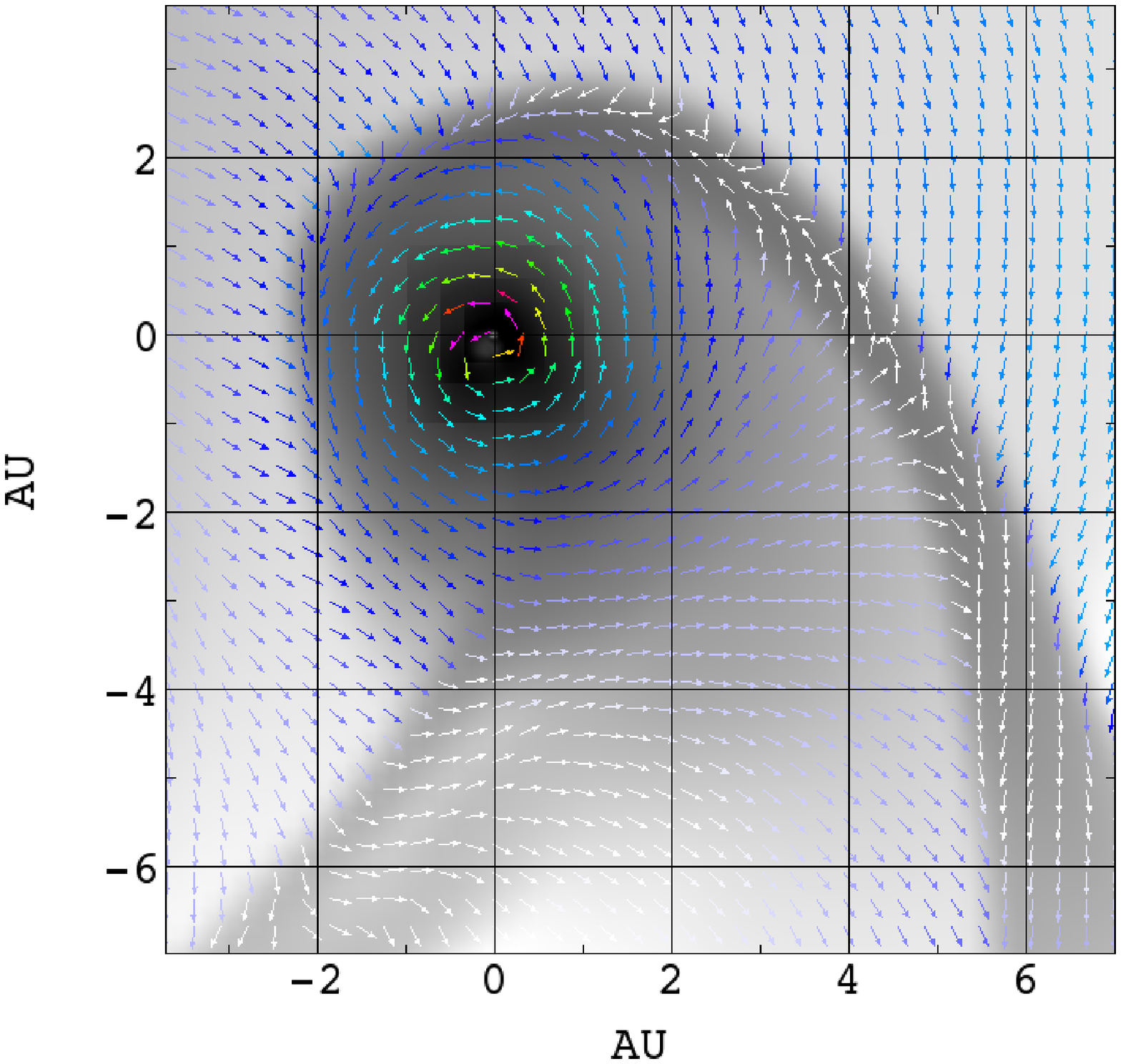}
      \includegraphics[width=.30 \textwidth,bb=125 175  580 660,clip=]{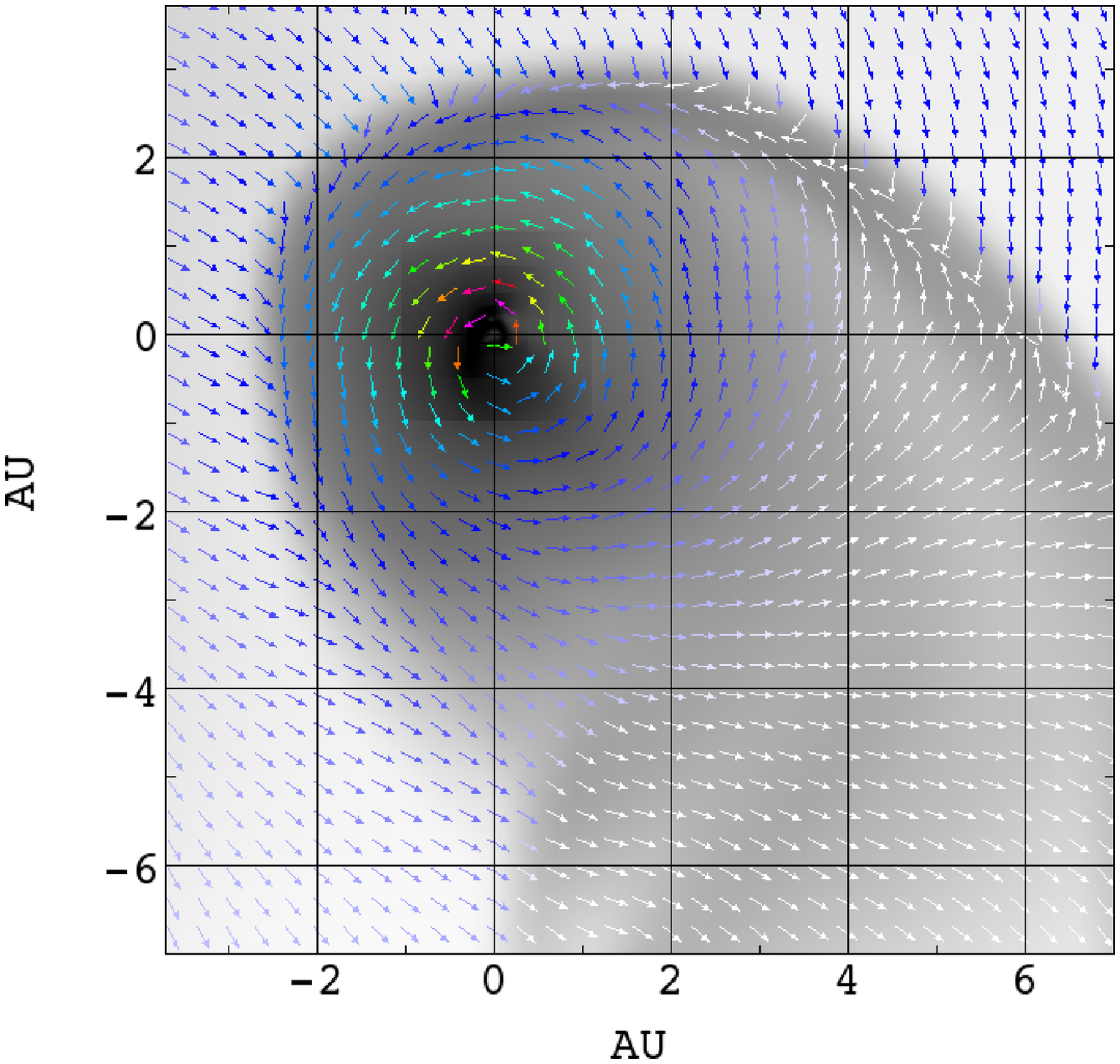}
      \includegraphics[width=.30 \textwidth,bb=125 175  580 660,clip=]{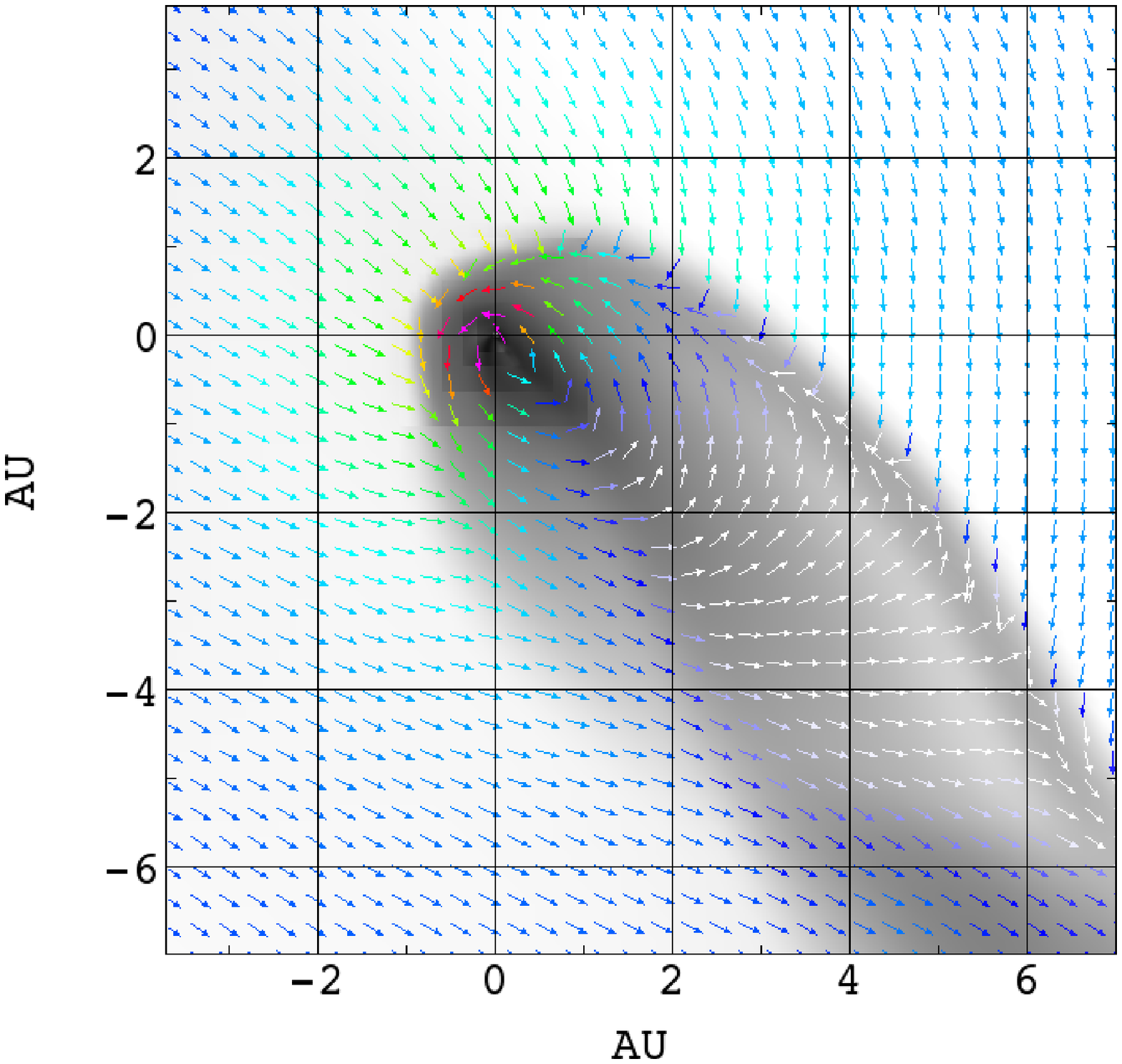} \\
~~~~~~~~~~~~~~Face-on, zoom in: \\
\vskip.10cm 
~~\,\includegraphics[width=.372\textwidth,bb= 20 175  515 655,clip=]{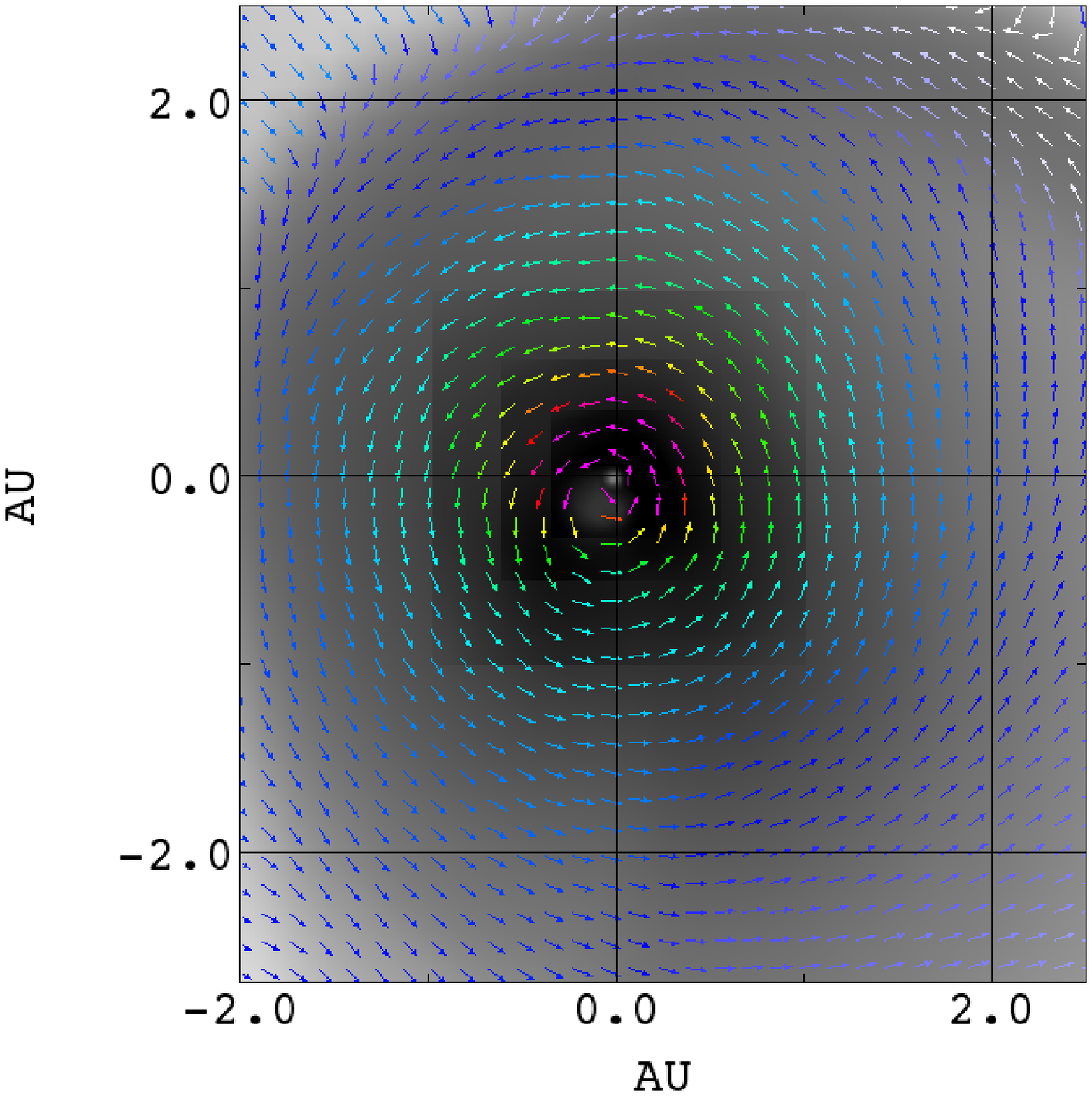}
    \includegraphics[width=.295\textwidth,bb=125 175  515 655,clip=]{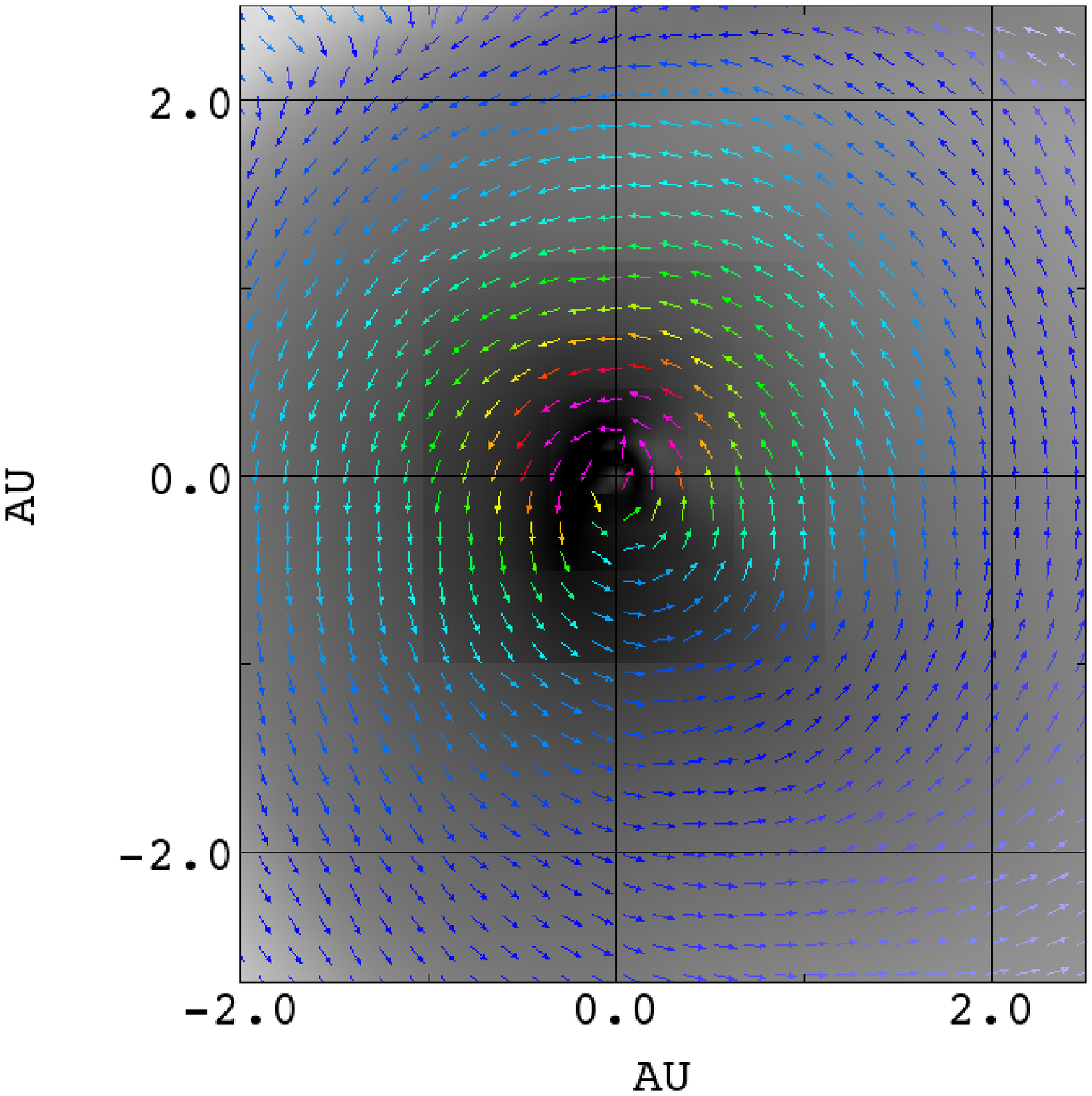}
    \includegraphics[width=.295\textwidth,bb=125 175  515 655,clip=]{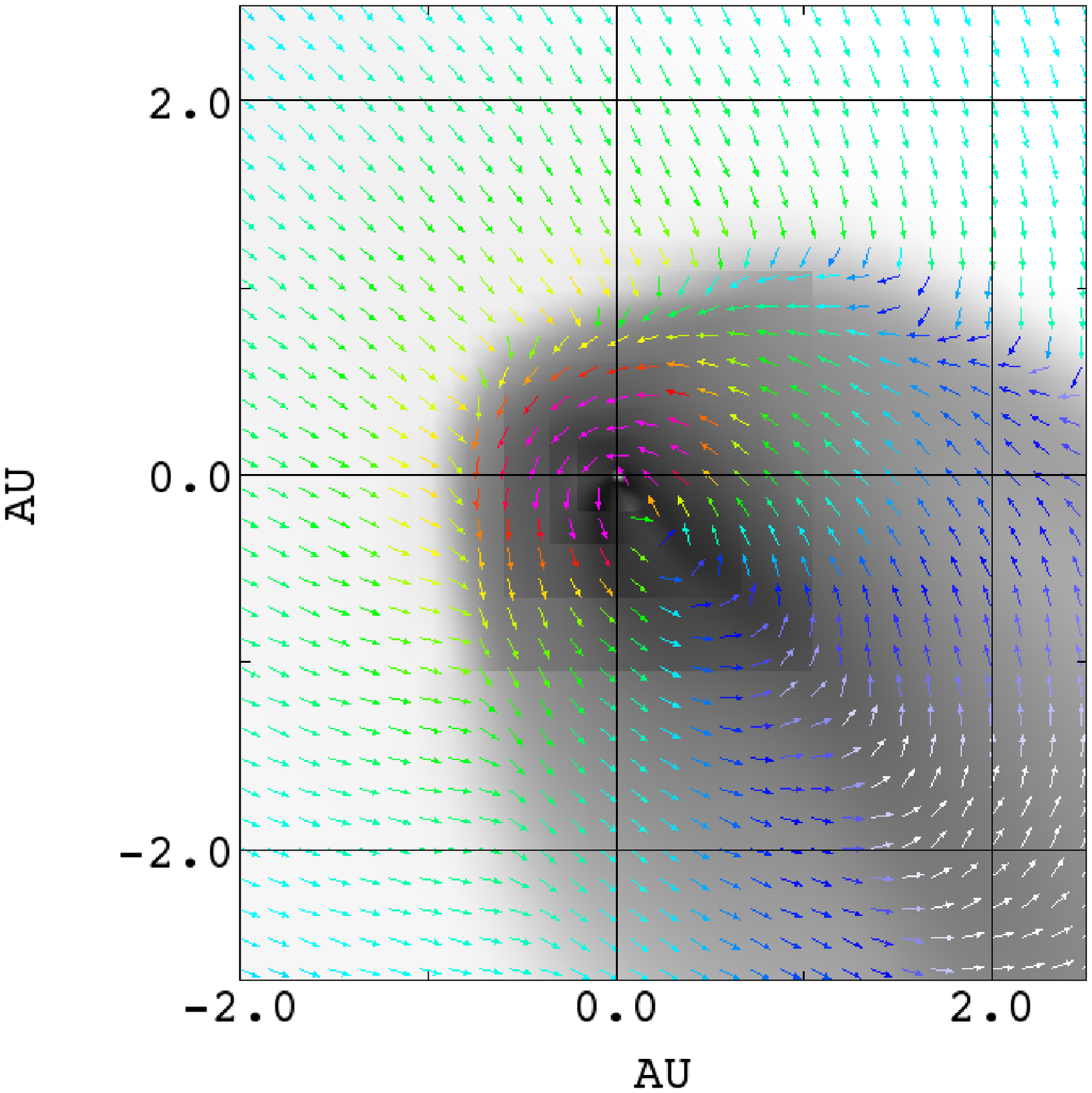} \\
~~~~~~~Edge-on: \\
\vskip.10cm
    \includegraphics[width=.365\textwidth,bb= 05 275  490 515,clip=]{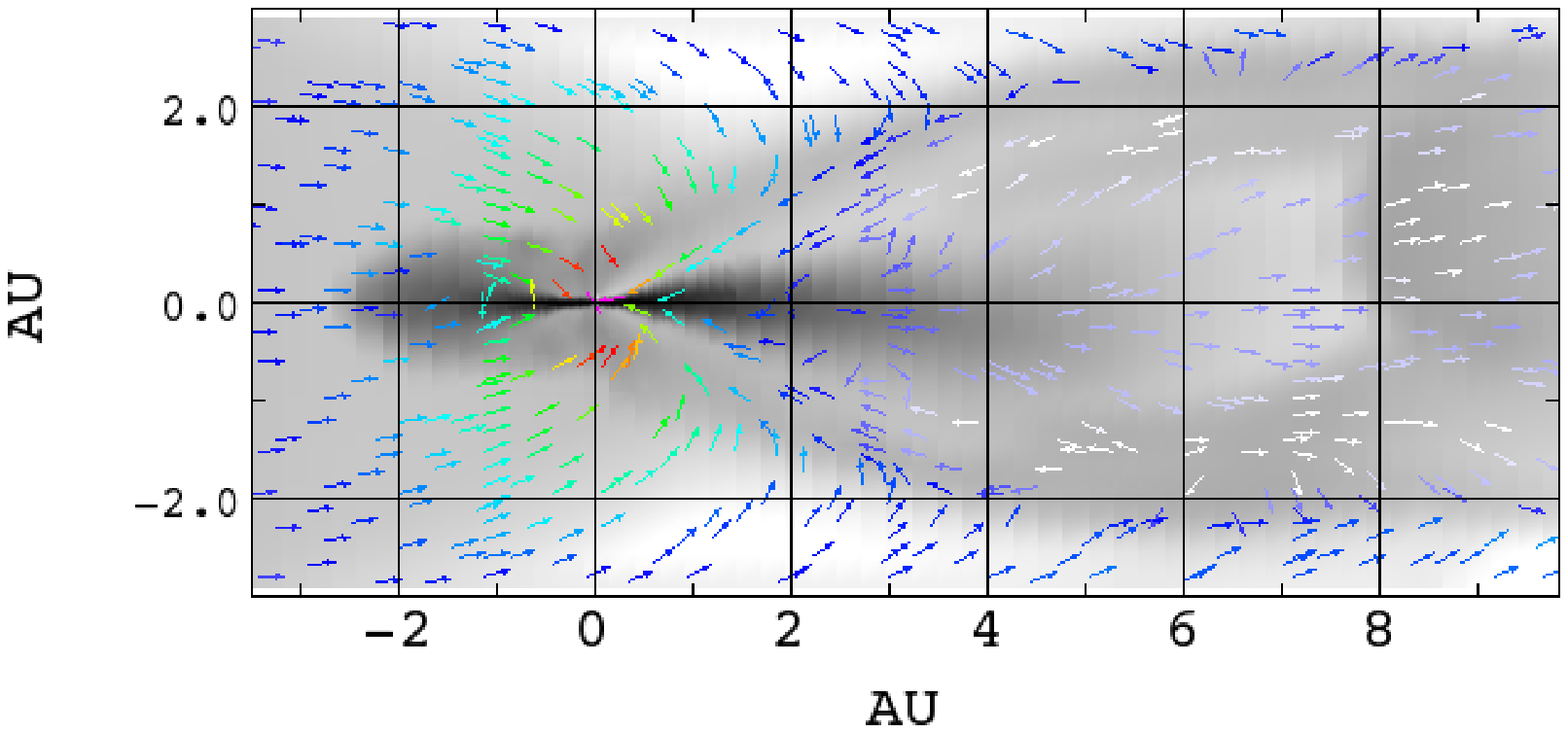}
    \includegraphics[width=.30 \textwidth,bb=088 275  490 515,clip=]{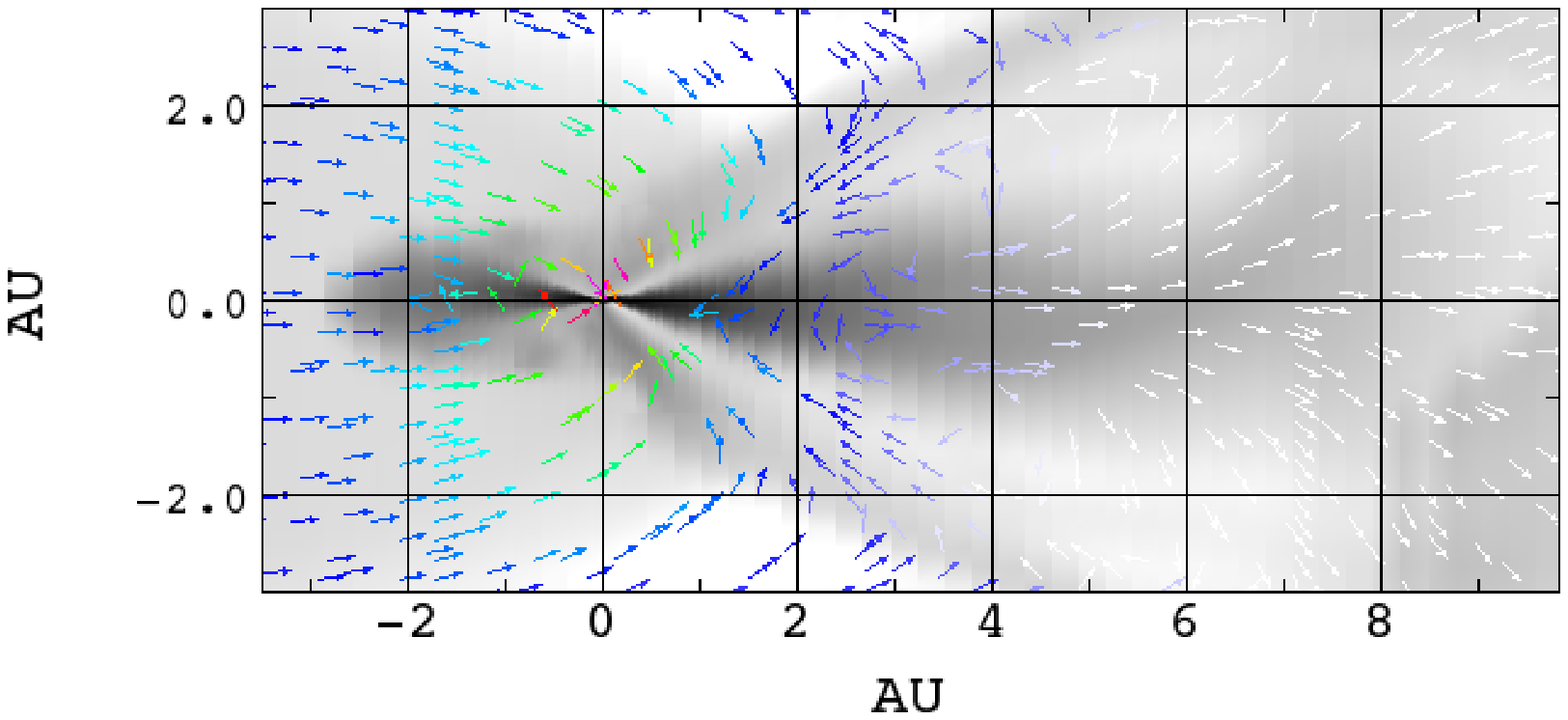}
    \includegraphics[width=.30 \textwidth,bb=088 275  490 515,clip=]{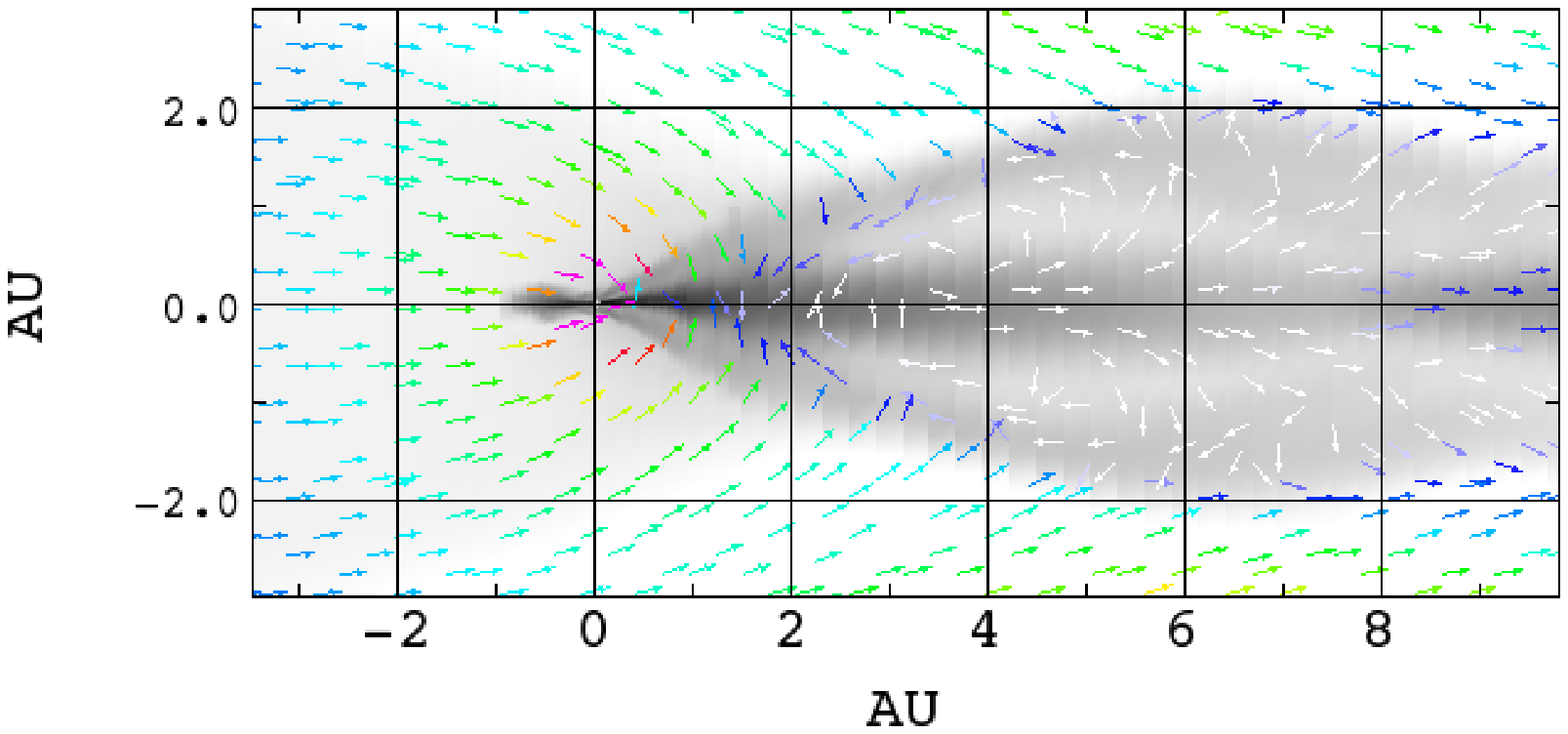}
  \caption{Logarithmic density gray-scale and velocity field maps
for the 10\,AU (left), 15\,AU (middle) and 20\,AU (right) models
at $t=\,$3\,orbits.  The primary is simulated to be to the left of
the maps. The
   secondary
is receding in bottom row panels due to its orbital
motion.
}
  \label{dens}
\end{figure*}

\subsection{Disk structure and formation}
\label{struc}

In order to present the main results of our study in Figure~\ref{dens}
we show logarithmic density gray-scale and velocity field maps of
our three simulations at $t=\,$3\,orbits. Rows~1 and~2 of
Figure~\ref{dens}
show slices though the orbital plane (at two different
scales), while the bottom row shows slices though a longitudinal
plane that intersects the secondary.  These density maps allow us
to see the large scale structure of the flows and to compare those
flows as the orbital separation is increased.

First we note the basic BHL converging flow which can be seen in
bottom row of Figure~\ref{dens}. The AGB wind can be seen streaming
in from the left and is subsequently focused towards the midplane
by the gravitational force of the secondary.  At the time of these
images the global flow is considerably more complicated than that
described by the simple BHL scenario, however, because of the presence
of fully formed accretion disks.
Thus, as Figure~\ref{dens} makes clear, accretion disks are created
in all our simulations.  These disks form, as expected, from the
BHL accretion stream that originates at the stagnation region (below)
downstream from the secondary.

\begin{figure}
\centering
  \includegraphics[width=.85\columnwidth]{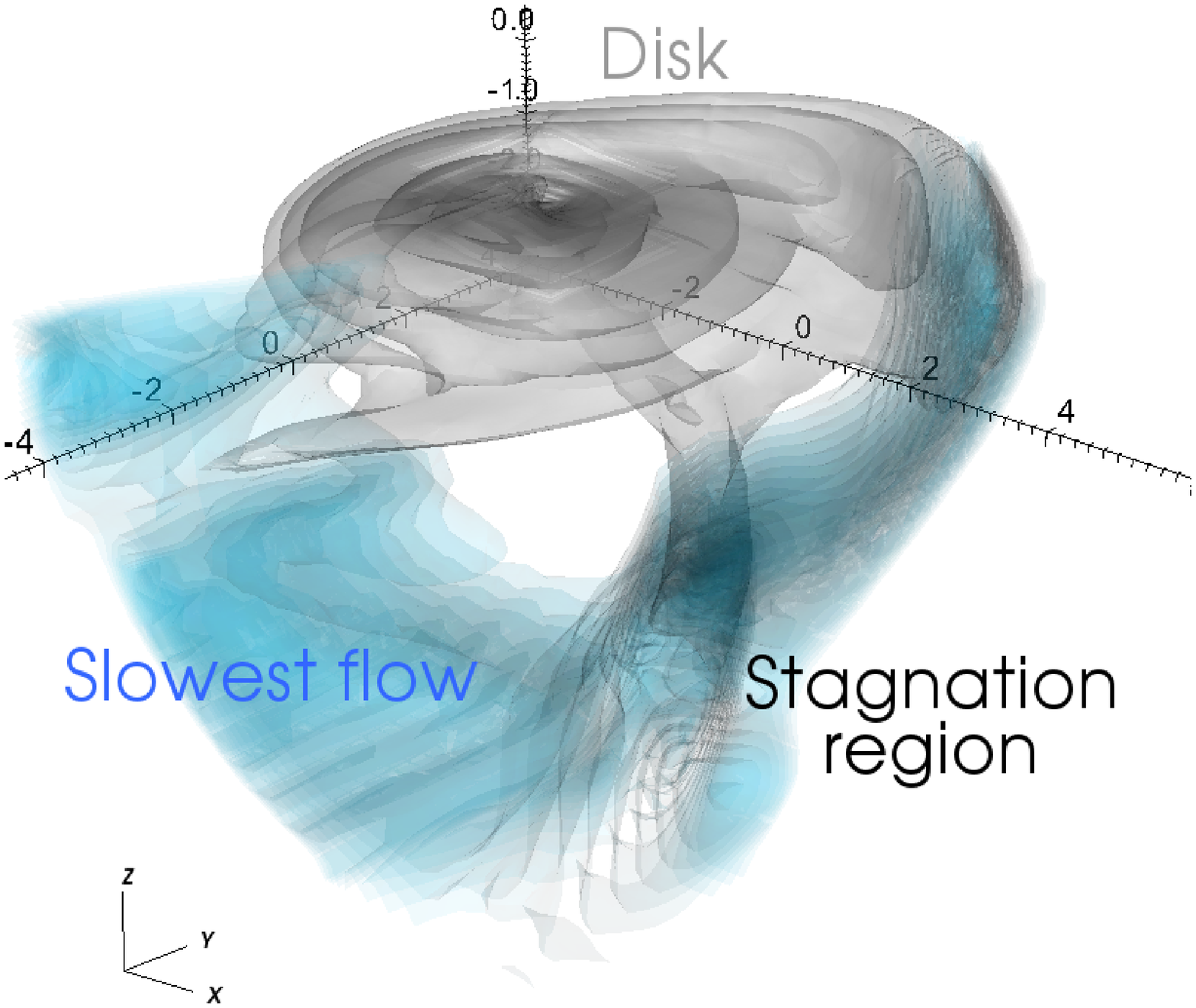}
  \includegraphics[width=.85\columnwidth]{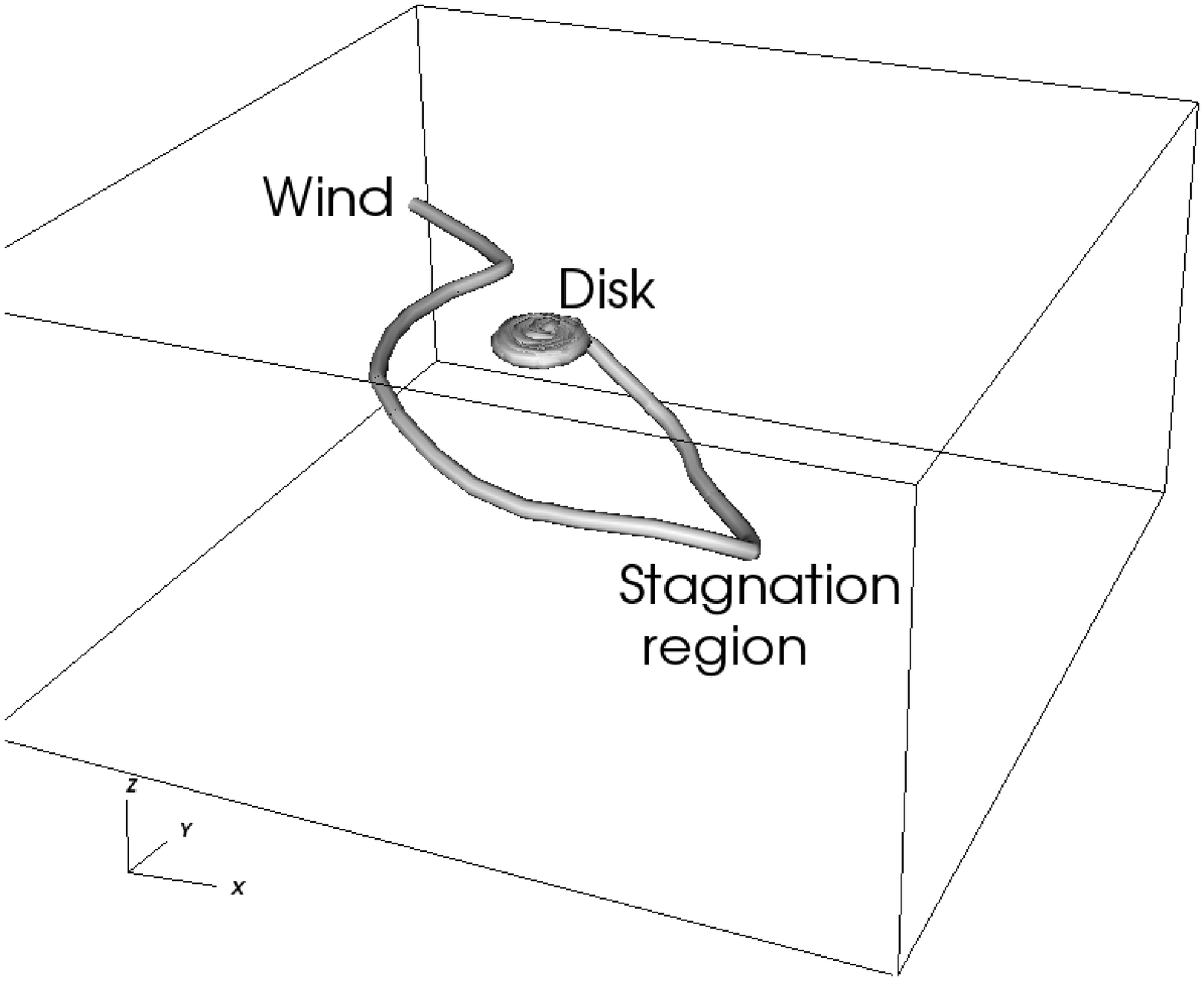} \\
  \includegraphics[width=.85\columnwidth]{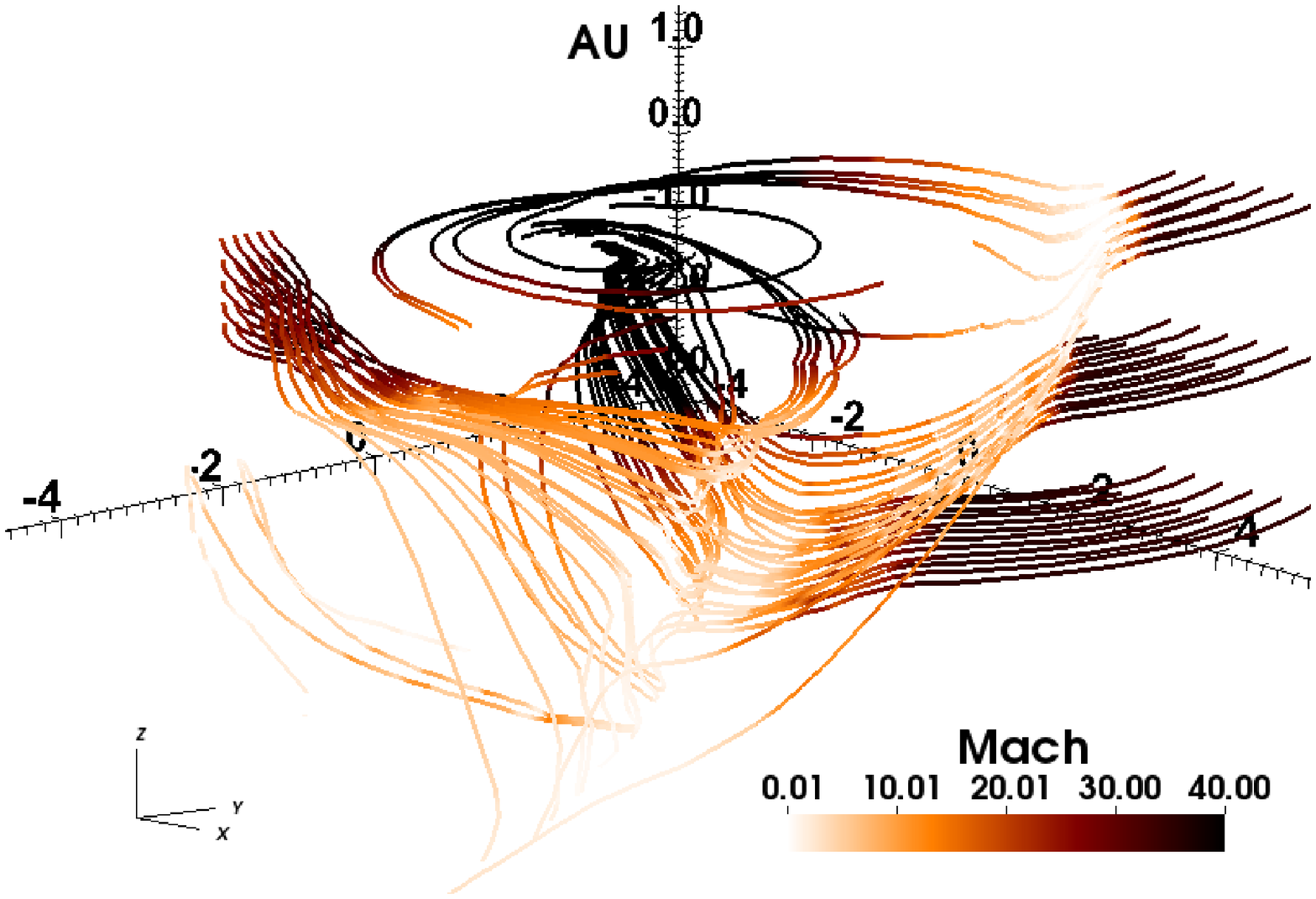} \\
\caption{Structure of the stagnation region for the 15\,AU case at
$t=\,$3\,orbits. Top: gray shows disk density iso-surfaces, while
blue shows velocity iso-surfaces of wind material with
speeds~$<$2\,km\,s$^{-1}$.  Middle: flow streamline that starts at
the boundary (wind), is deflected by the secondary's gravitational field,
reaches the stagnation region and is finally accreted onto the
disk. Bottom: flow streamline field with a color scale in Mach units.}
  \label{stag}
\end{figure}

We provide a better illustration of the stagnation region in
Figure~\ref{stag} where the top panel shows iso-surfaces
of the disk's density and slow material's velocity. The middle
panel shows one wind flow streamline which starts at the
boundary, is
deflected by the secondary's gravity field, reaches the stagnation region then,
and is finally accreted onto the disk. In the bottom panel we show
several flow streamlines from a different perspective using
a velocity color table that clearly shows the stagnation region structure. 
We note the appearance of nested flow surfaces,
whose areas increase with distance from the stagnation
region. Thus the accretion column resembles a strong vortex tube 
	--instreaming material flowing along helical paths towards the 
	secondary star--
and our simulations are the first to capture this aspect of BHL 
flows in binaries.

The density maps in both the equatorial and perpendicular planes
(Figure~\ref{dens}) also reveal the presence of bow shocks around the accretion disks.
Such shocks also form in the classic BHL flow however once the
disks emerge, incident AGB wind must stream around them creating
more complex post-shock flow patterns.  The velocity vectors in the upper row panels
of Figure~\ref{dens} are of particular interest, showing the
redirection of high density flow behind the bow shock and its
interaction with the counterstreaming flow in the accretion disk.

\begin{figure*}
\centering
  \includegraphics[width=.40\textwidth,bb= 75 170 596 685 ,clip=]{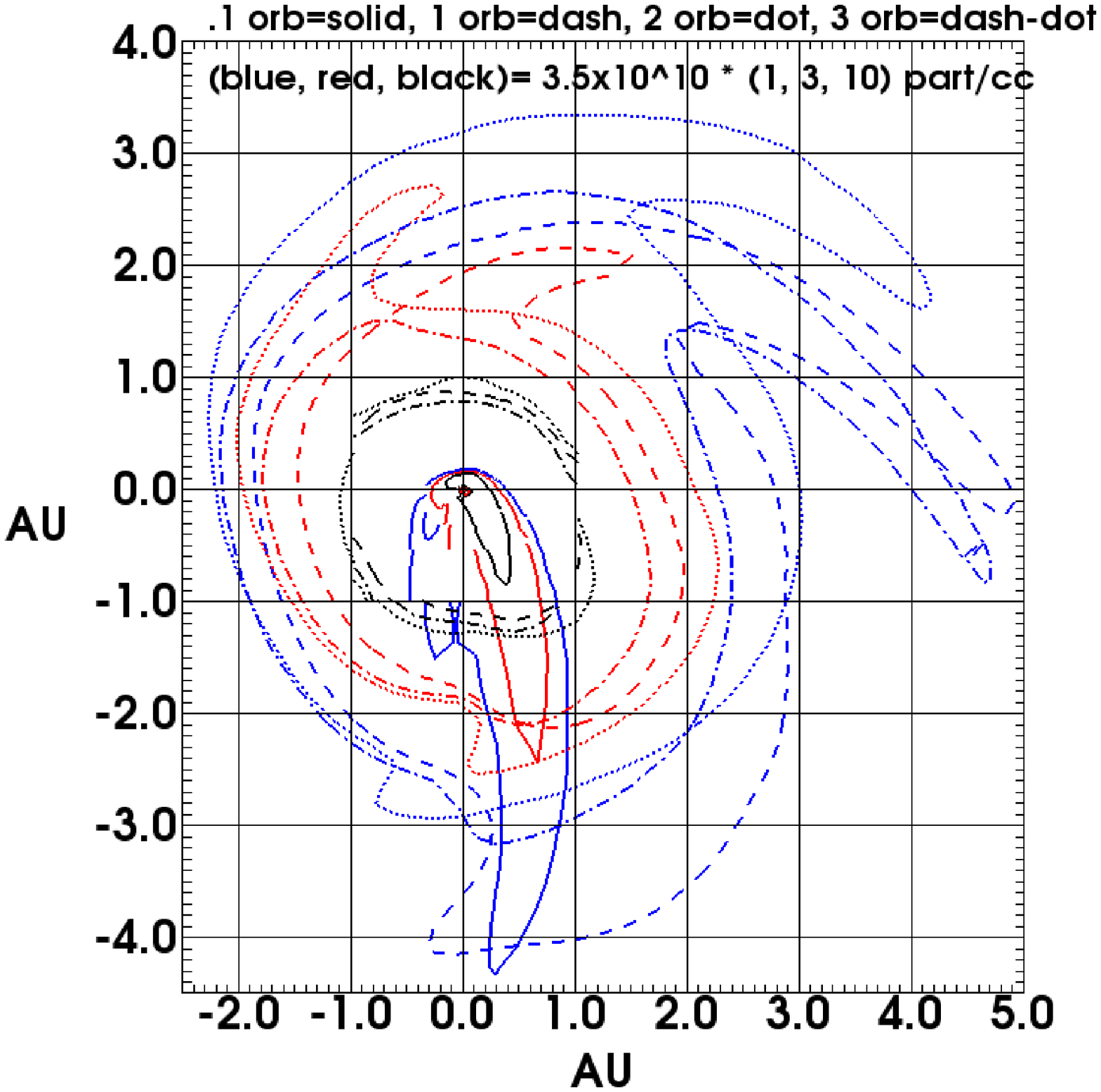} 
  \includegraphics[width=.40\textwidth,bb= 55 120 596 521 ,clip=]{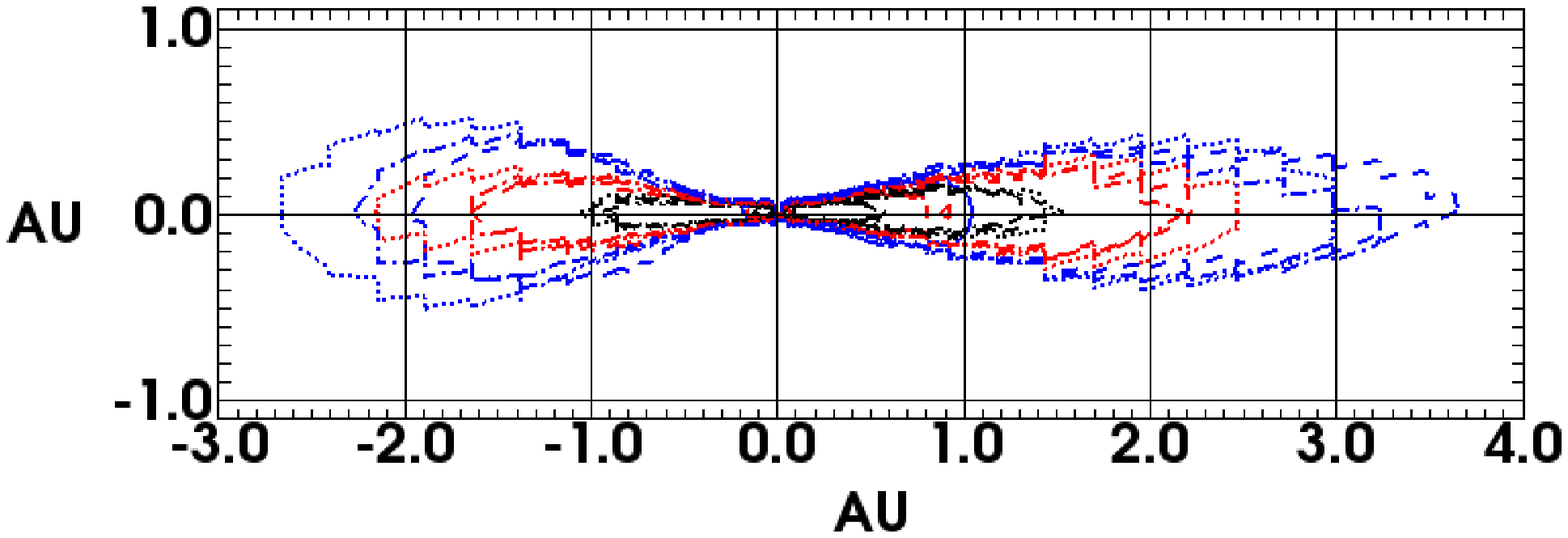} \\
  \includegraphics[width=.40\textwidth,bb= 70  185 570  650,clip=]{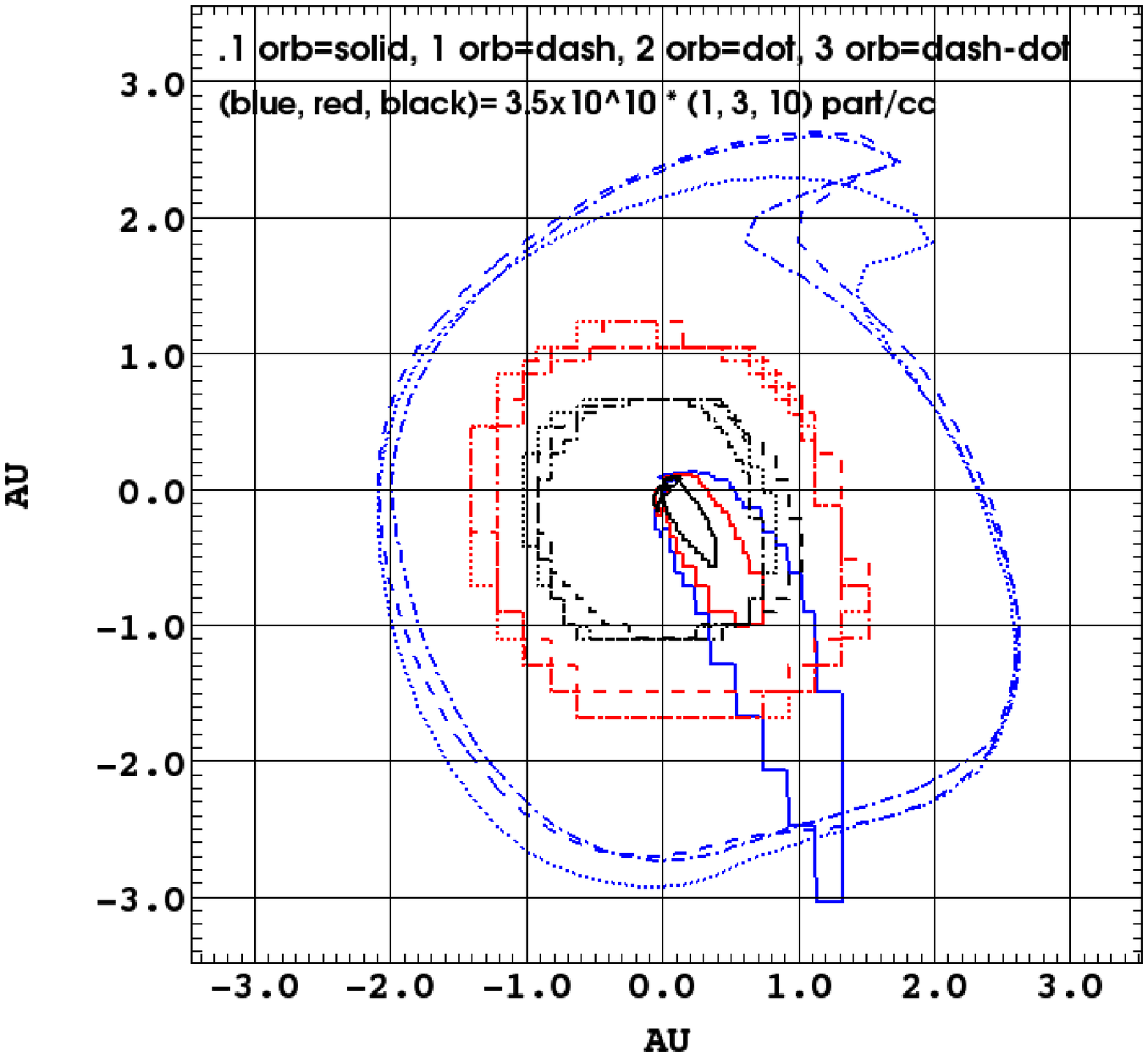} 
  \includegraphics[width=.40\textwidth,bb= 85  155 590  480,clip=]{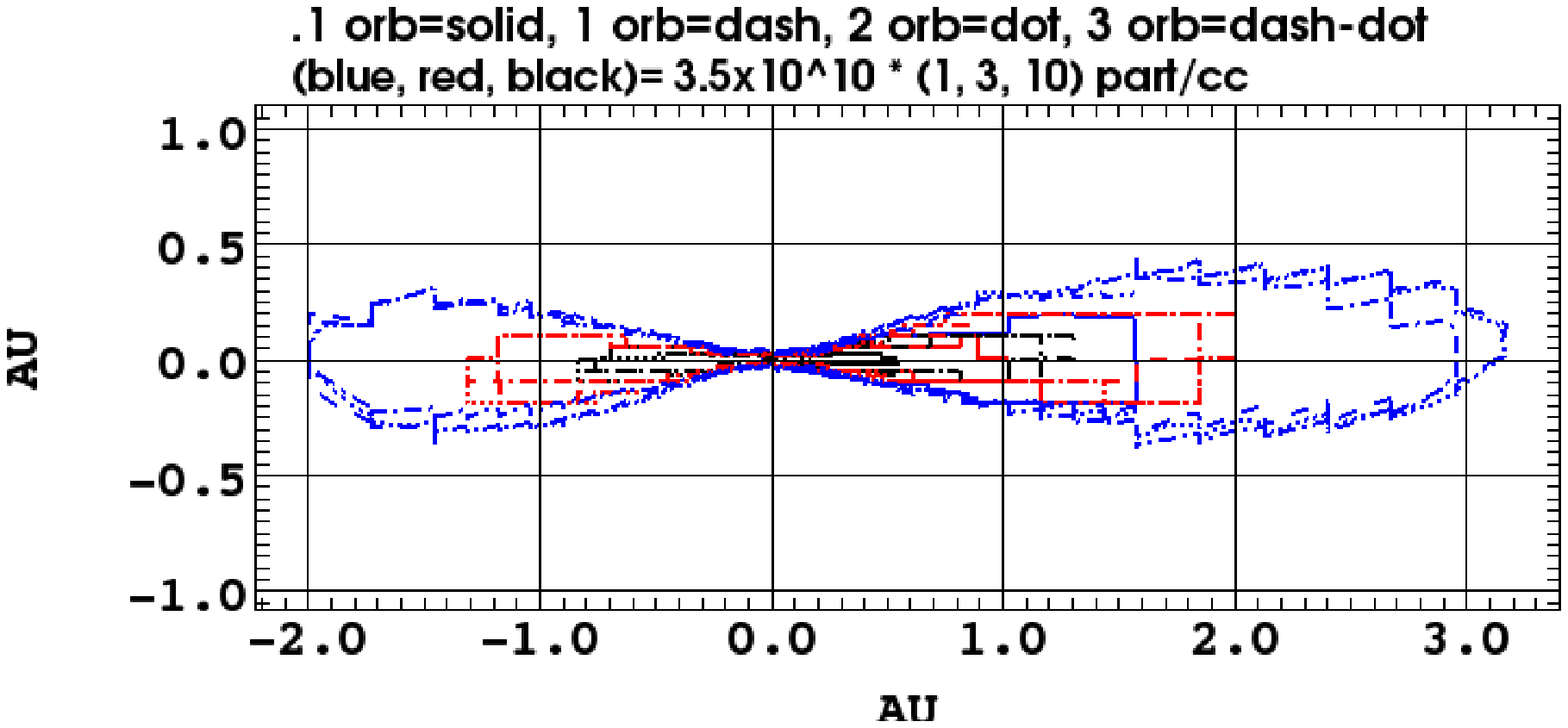} \\ 
  \includegraphics[width=.40\textwidth,bb= 60   45 730  655,clip=]{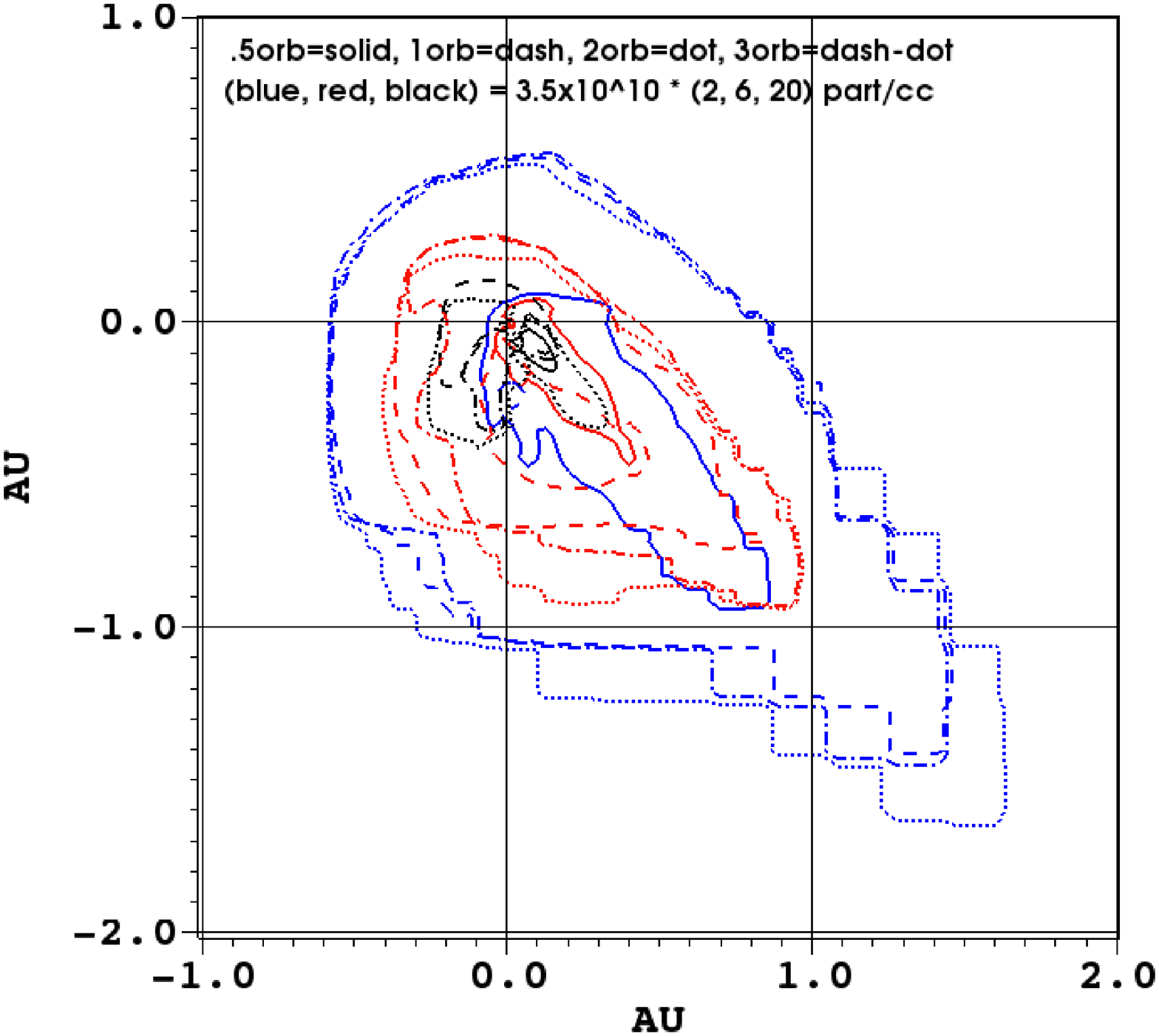} 
  \includegraphics[width=.40\textwidth,bb= 70 -410 1140 420,clip=]{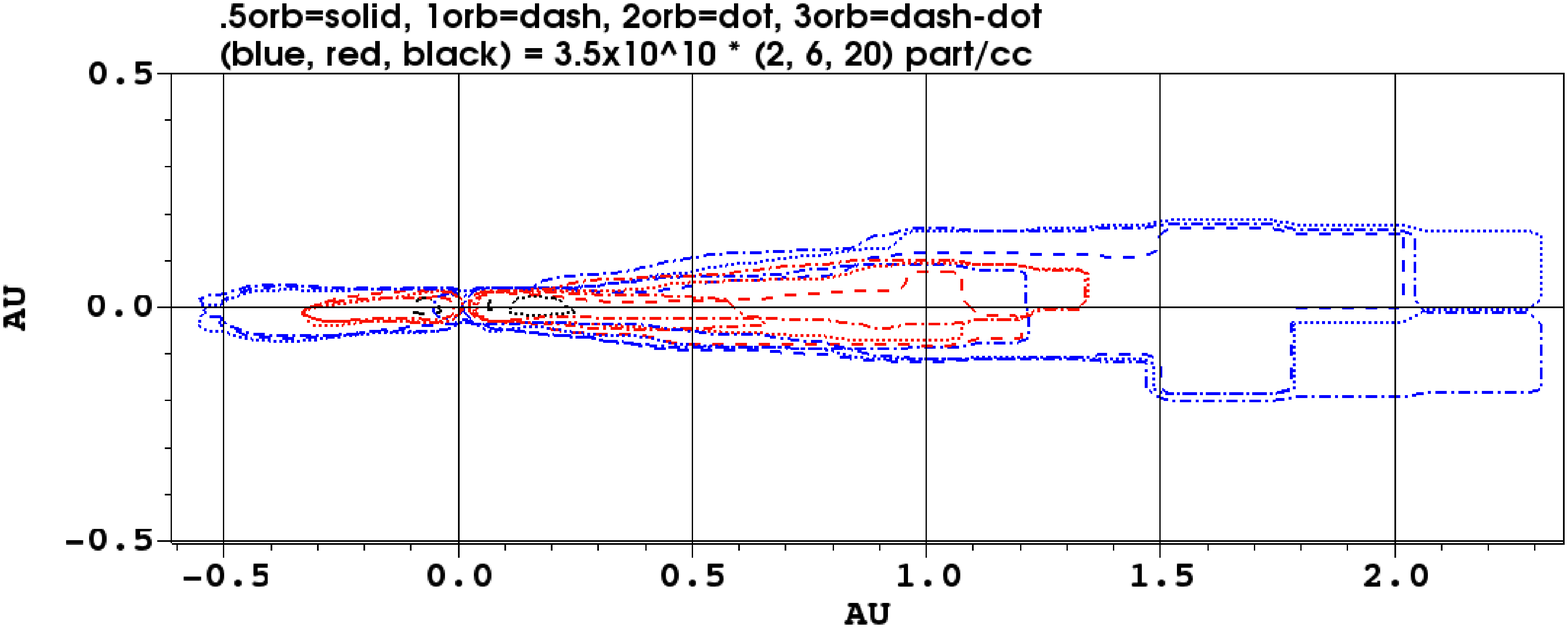} \\ 
\caption{Disk density structure at 4 times. Left: face-on view where
the wind enters the grid from the top left corner at an angle close
to 45$^{\circ}$ and the secondary orbital motion is upwards.  Right:
edge-on view where the wind enters the grid horizontally from the
left side and the secondary is receding due to its orbital motion.
The primary is simulated to be to the left of all maps.  Top, middle
and bottom rows correspond to binary separations of 10, 15~and
20\,AU, respectively.}
  \label{struct}
\end{figure*}

To understand the structure and evolution of the disks more explicitly,
in Figure~\ref{struct} we show logarithmic density contours of the
disk in slices though the orbital and perpendicular planes. Each
panel has 12 contours arranged by line type and color: blue, red
and black represent densities of (3.5, 10.5, 
35) $\times$10$^{10}$\,part\,cm$^{-3}$, respectively.  Dashed, dotted
and dashed-dotted~lines represent densities at $t=\,$1, 2~and
3\,orbits, respectively.

The density contour maps for different orbital separations are
mutually consistent in terms of evolutionary stages but are
correspondingly different with respect to  the  time scales  and
some features  of these phases.  Comparing contours of the same
color (Figure~\ref{struct}) we see a given iso-density structure at
different evolutionary times.  These show that once the disk forms
further structural changes are mild. Thus in all cases the disks
begin ($t =\,$0.1) from long elliptical structures that can be
identified as the curved BHL accretion stream. By ($t =\,$1.0)
the inner regions of the disk have formed and we see well defined
inner black contours which do not change appreciably with time.
The middle and outer contours in all three case (red and blue) do
show some evolution after $t=\,$1.0. However, these changes tend to be
higher order in morphology, whereas overall shape (circular in the
$a=10$ and $15$\,AU case and elliptical in the $a=20$\,AU case) and
radii do not change.

Comparing the solid lies in left panels we find, in agreement with
the discussion in section~\ref{bondi}, that the angle between the
accretion flow and the wind decreases with 
increasing orbital separation,
$a$. This angle is close
to zero already in the 20\,AU model. Thus, as expected, the solution
(\ref{par3}) converges to the BHL one for large separations.

We note that given our choice of $\gamma = 1$, the accretion disks
which form in our models are thin which is consistent with a low
gas temperature and sound speed ($h/R \sim c/\Omega)$.  The disks
orbits are also not circular.  We see orbits with eccentricities
$e > 0$ in all cases, however the magnitude of the average eccentricity
increases with increasing orbital separation.  And while our
disks are thin, we do see flared vertical structures that are also
not symmetric in the ($r,z$) plane of the disk but larger on their
downstream side relative to the wind. The effect of the wind on
the disk is also apparent in that the steepest density gradients
are located where the disks impinge on the incoming stellar flow
(top left region in left panels and left region in right panels,
Figure~\ref{struct}).
We find that both the disks' outer radius, $r_d$, and edge height
are inversely proportional to $a$. In section~\ref{orbi} we show
that the shape of the gas parcel orbits in the disk is a function
of $a$ as well. We note that the disks' inclination angles are
generally small, meaning that the disks are close to perpendicular to
the orbital plane. However, small time
variation in the flow occur even after the disks settle into a
quasi-steady state.

\subsection{Disk mass}
\label{disk_mass}

\begin{figure*}
\centering
  \includegraphics[width=.030\textwidth,bb= 050 220 90 585, clip=]{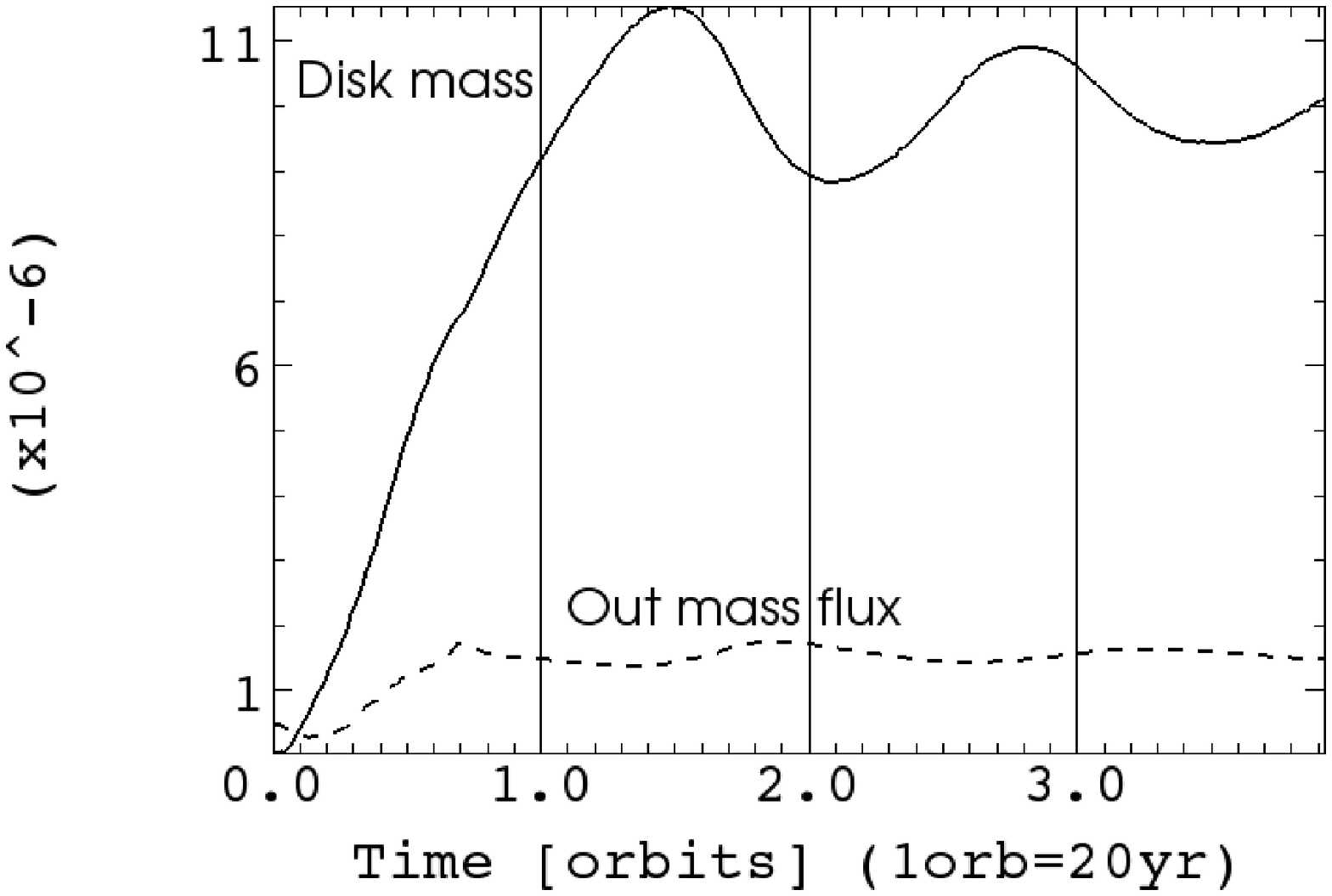}
  \includegraphics[width=.480\textwidth,bb= 050 220 580 585, clip=]{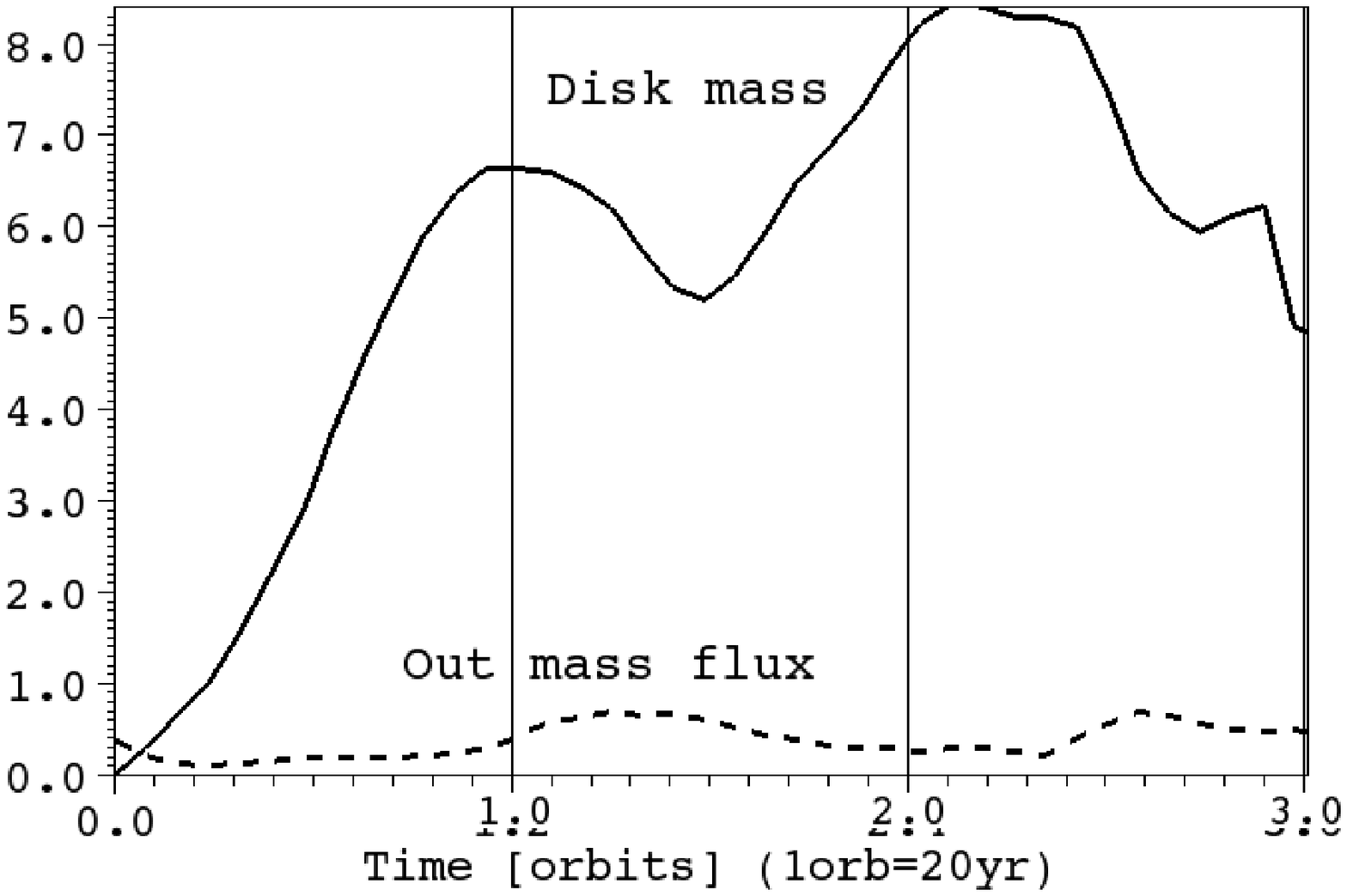}
  \includegraphics[width=.45\textwidth,bb= 090 210 580 570, clip=]{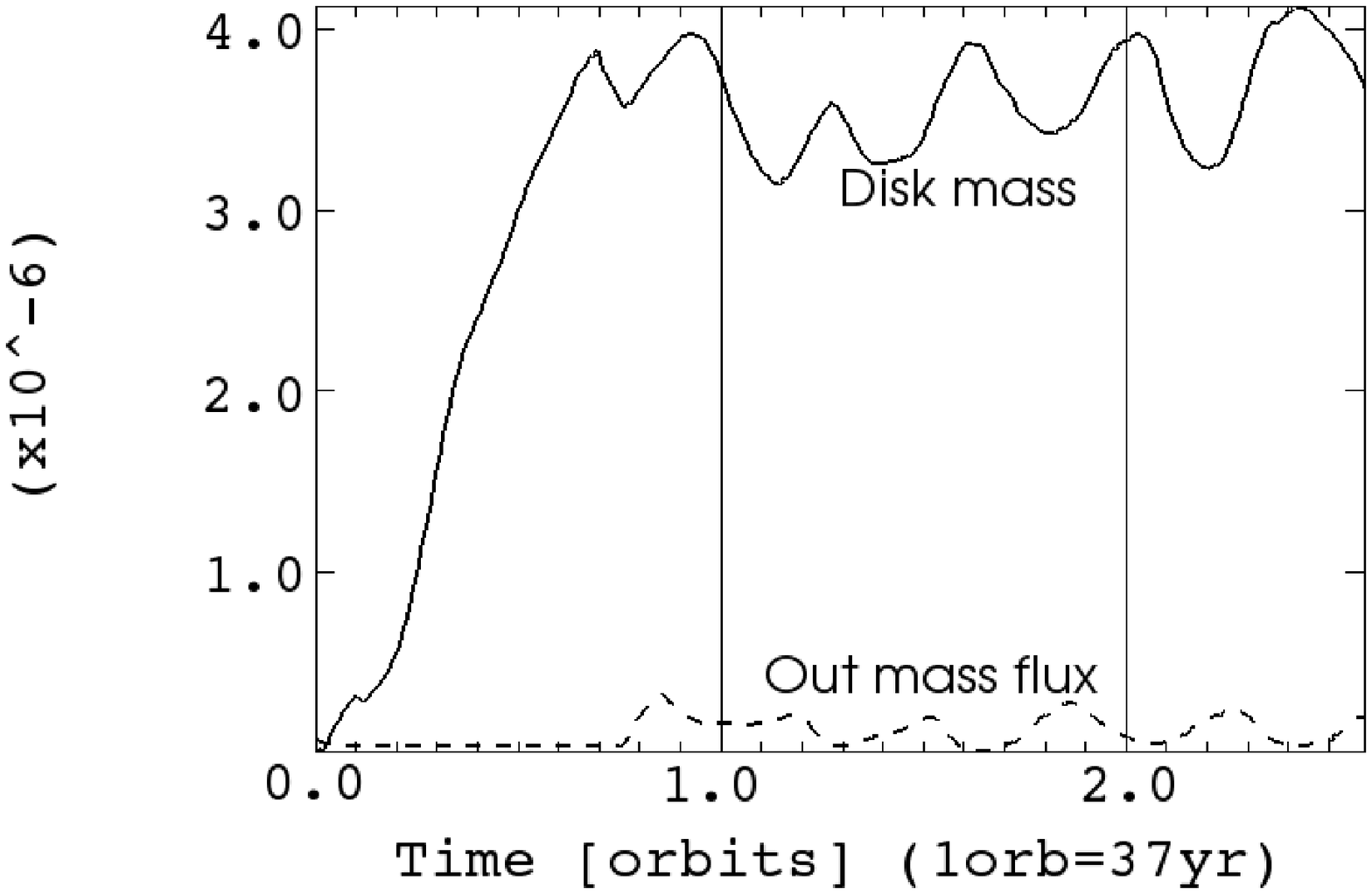}\\
\textit{(a) $a=\,$10\,AU \hskip6cm (b) $a=\,$15\,AU} \\
  \vskip.3cm
  \includegraphics[width=.505\textwidth,bb= 090 210 580 570,
  clip=]{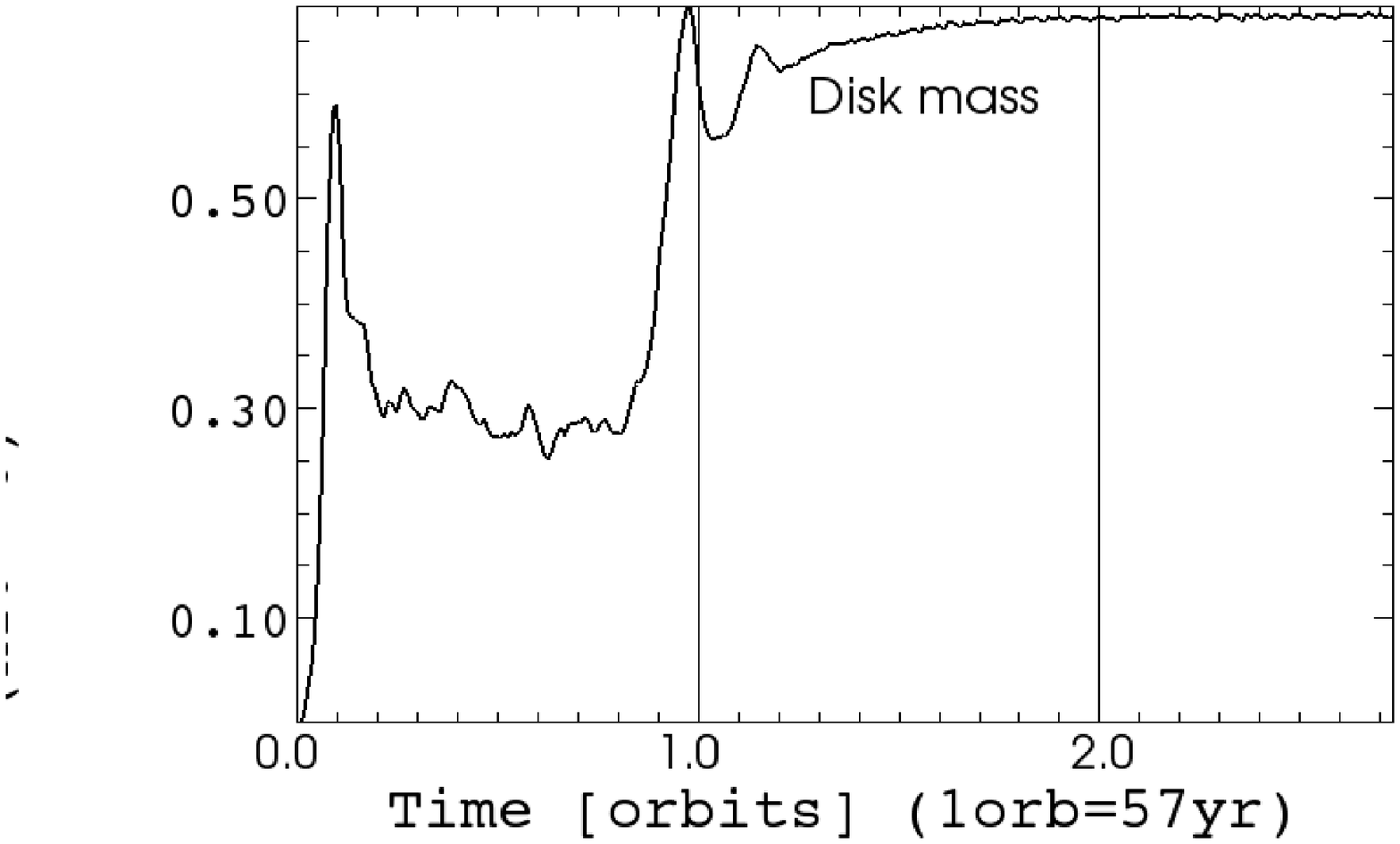}\\
\textit{(c) $a=\,$20\,AU} \\
\caption{Disk mass as a function of time [M$_{\odot}$] (solid) 
and flux of gas that leaves the grid [M$_{\odot}\,$yr$^{-1}$]
(dashed). 
}
  \label{mass}
\end{figure*}

In Figure~\ref{mass} we show profiles of the disks' mass,
$M_d(t)$ (solid
lines), as a function of both time and the binary separation.  These are
calculated by summing the mass of grid cells containing bound gas
(those having a gravitational energy greater than the kinetic one).

In the 10\,AU case we see that the disk mass increases during the
first orbit and then saturates at about 7$\times$10$^{-6}$\,\ms.
The disk mass then oscillates with a period of $\sim\,$1\,orbit and
an amplitude of about 1$\times$10$^{-6}$\,\ms.

To understand the origin of these disk mass oscillations, we calculate
the gas flux leaving the grid though both the $+x$ and $-y$ boundaries
(see Figure~\ref{diag})
in \my\ (dashed lines, Figure~\ref{mass}).  The position and shape
of the gradients in $M_d(t)$  compared with 
those of
the mass fluxes indicate
that the disk's mass decreases with increasing outbound flux and
vice versa.  This suggests  the wind's post bow shock ram pressure
mediates the disk mass;  the wind appears able to strip a small
fraction of disk's gas and liberate it from the gravitational pull
of the companion.
This is likely to happen at the outer parts of the
disk where the ram pressures of the wind and the disk are comparable.
Thus the wind-disk interaction induces perturbations on the orbits
of the disk material, which may be sufficiently strong to overcome the
gravitational potential of the companion. Some disk gas is pushed outside
the computational domain then.

Moving to comparisons of simulations with different $a$, we see
that  the disk mass in  the 15\,AU model reaches its maximum value
of about 4$\times\,$10$^{-6}$\,\ms\ in $\sim\,$0.7\,orbits (earlier
than the $a=10$ AU case). Once again we see disk mass oscillations
with  amplitudes of about 1$\times\,$10$^{-6}$\,\ms\ and a frequency
of $\sim\,$0.3\,orbits.  Note that these oscillations have a smaller
amplitude and period than for the 10\,AU case.  This would be
consistent with our wind stripping interpretation as the wind's ram
pressure is proportional to $a^{-2}\,{\bf x}_s^{-1}$, where ${\bf
x}_s^{-1}$ is the distance between the secondary and the center of
mass.

The disk mass profile of the 20\,AU model is quite different. While
an initial disk forms with a mass $\sim\,$6.5$\times$10$^{-7}$\,\ms,
a sharp growth in mass is seen at $t \approx\,$0.85\,orbits. The
increase occurs in less than $0.1$ of an orbit ($t =5.5\,yr$) and the
disk mass then reaches a maximum value of
$\sim\,$6.5$\times$10$^{-7}$\,\ms\
after a brief period of relaxation. Note that in this case we do
not see any disk mass oscillations. This behavior is 
also 
consistent with our wind stripping interpretation as the model wind ram pressure
is the weakest at 20\,AU 
away from the primary.

\subsection{Disk orbits} \label{orbi}
\begin{figure*}
\centering
  \includegraphics[width=.14\textwidth,bb= 35 470 118 645,clip=]{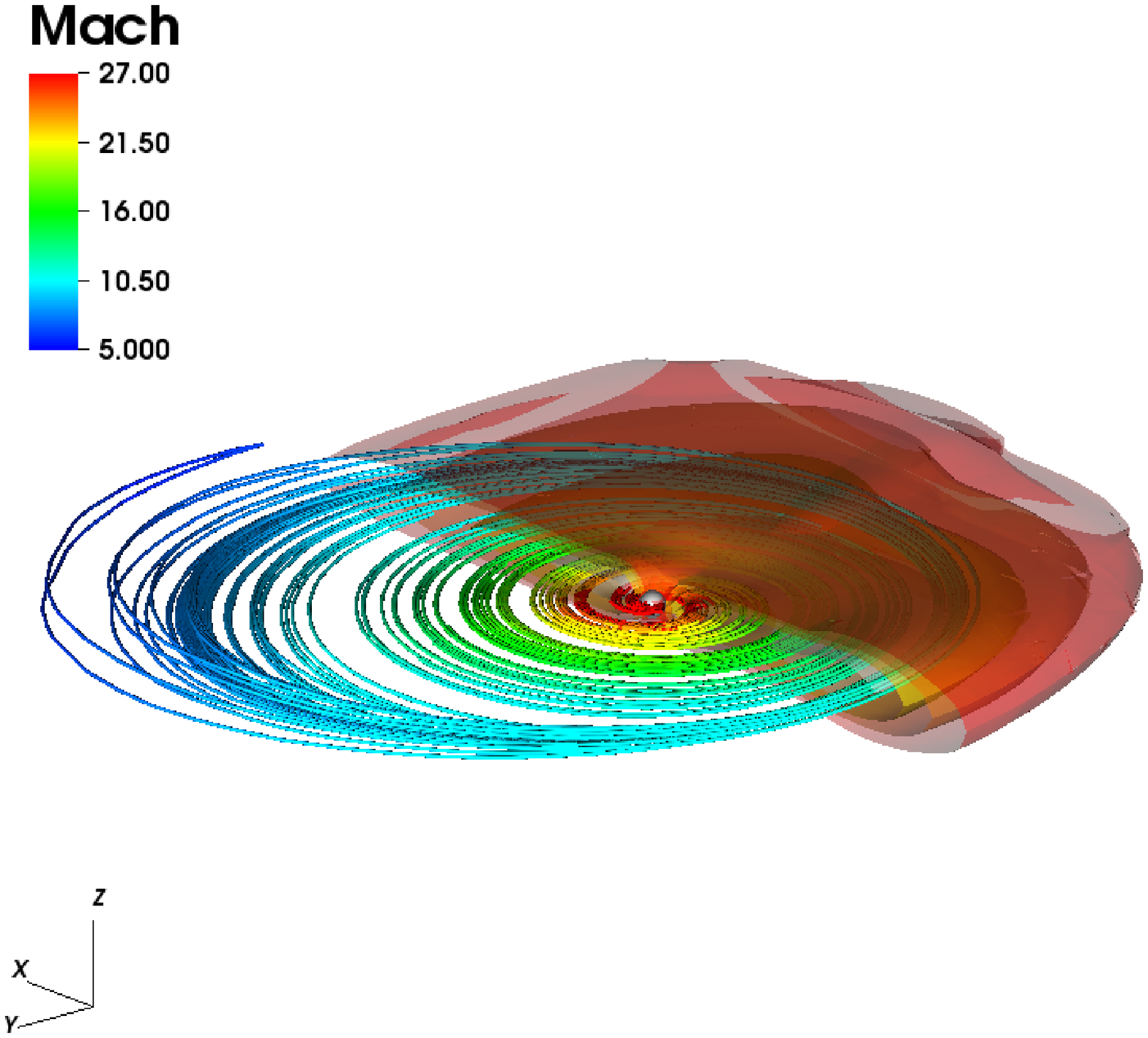}
  \includegraphics[width=.845\textwidth,bb= 40 290 560 470,clip=]{cool2.ps} \\
\vskip.3cm
  \includegraphics[width=.32\textwidth]{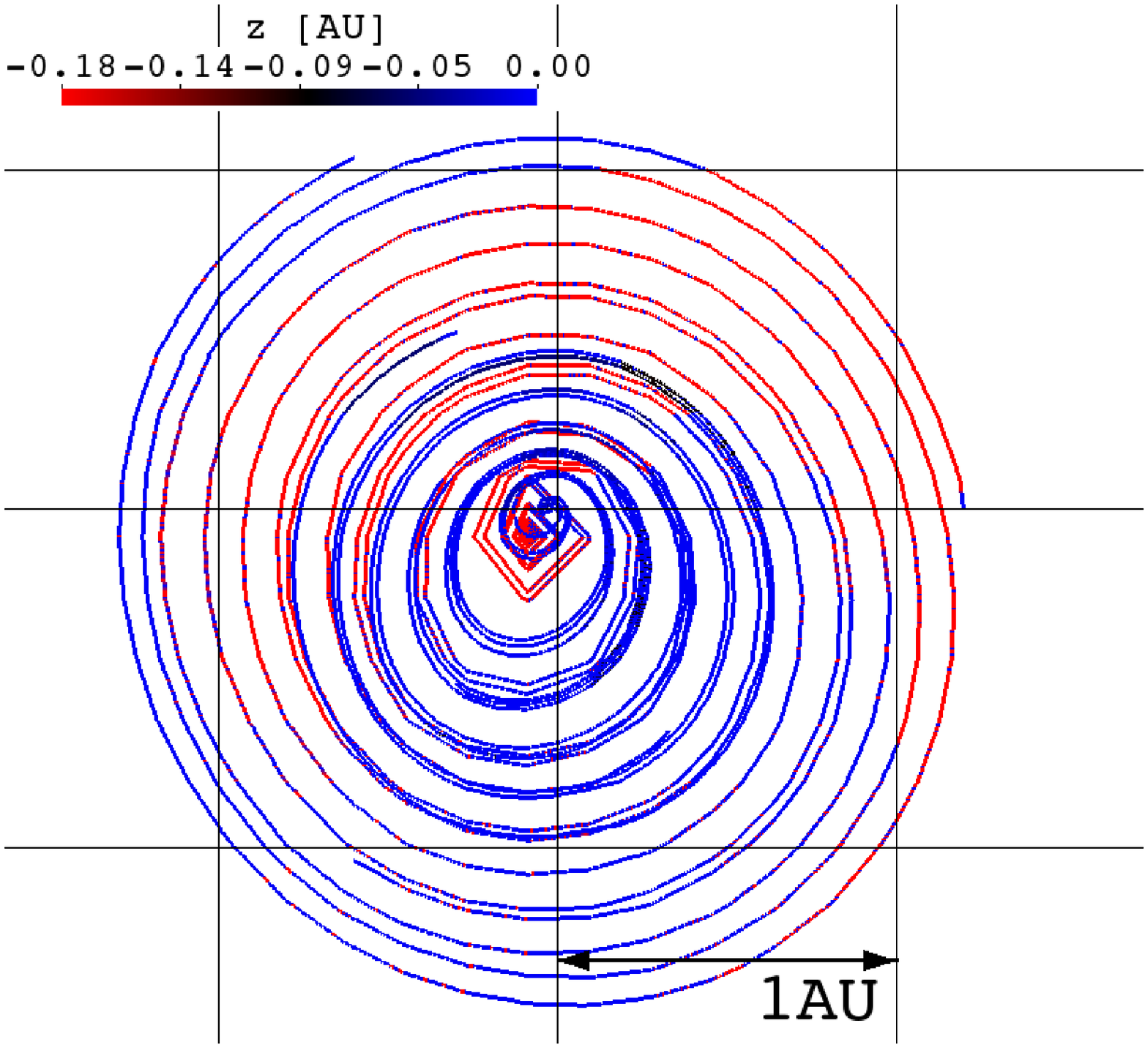}
  \includegraphics[width=.32\textwidth]{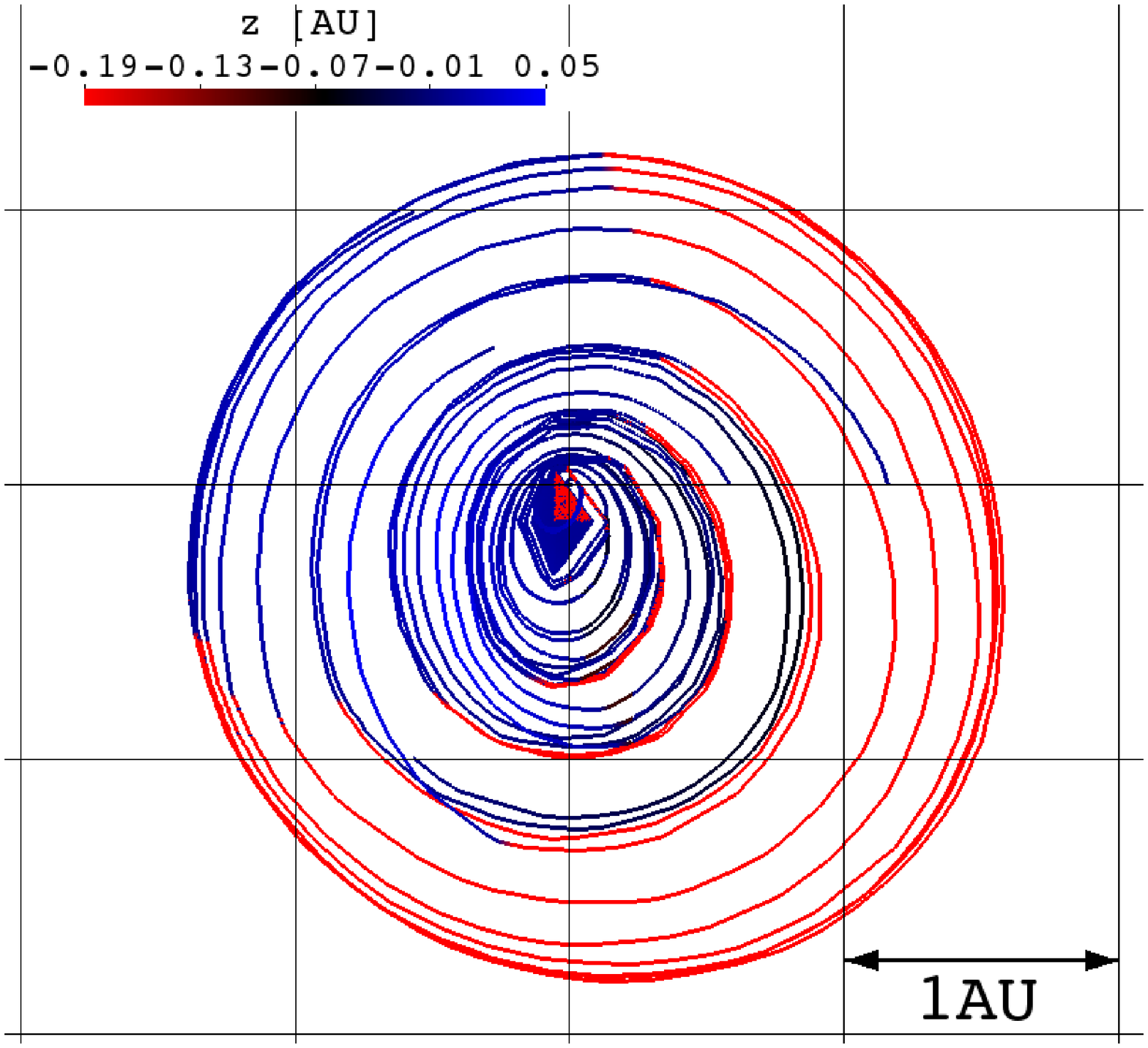}
  \includegraphics[width=.32\textwidth]{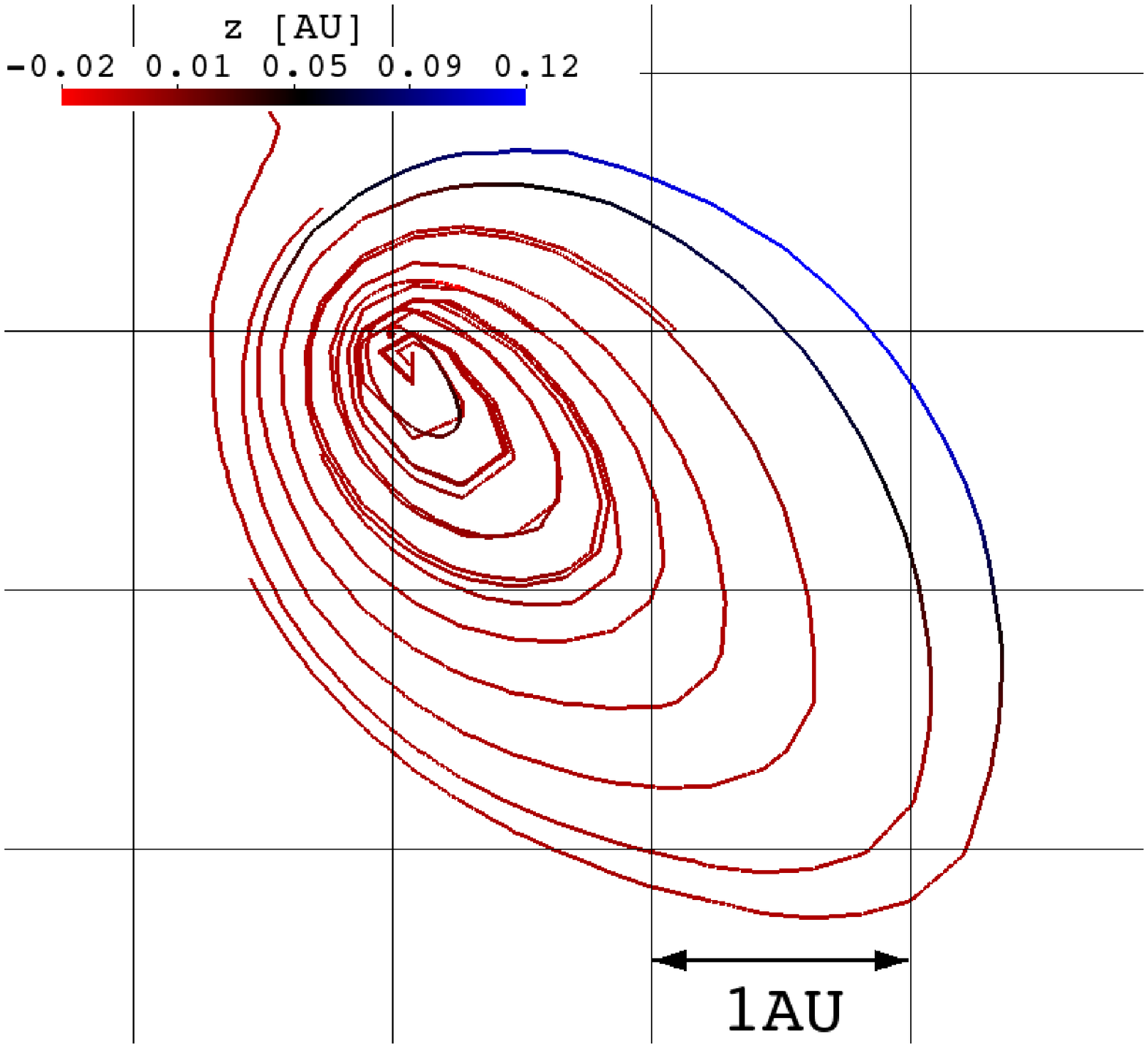}
\caption{Top: 3-D disk gas orbit streamlines and sliced density
contours (red) for the 10\,AU case at $t=\,$2\,orbits. The central
white small sphere is the secondary. Streamline colors
denote orbital speed in Mach units.  The wind enters the grid towards
the image. Bottom: Disk gas orbit streamlines at $t=\,$3\,orbits
corresponding to the 10, 15 and 20\,AU cases, from left to right.
The wind enters the grid from the top
left corner moving towards the bottom right.  Colors denote distance
perpendicular to the orbital plane in AU and indicate the orbits inclination.}
  \label{orb}
\end{figure*}

An interesting result from our simulations is that the 
shape of the disk gas orbits significantly depends on $a$. We
show this relation using the disk gas streamline maps in Figure~\ref{orb},
where the top panel presents a 3-D map of the disk gas velocity
distribution in colors for the 10\,AU model, along with sliced (half
the disk) density contours in translucent red color. The lines show
that the gas is supersonic, the orbits have low eccentricities,
$e$, and the velocity decreases with distance from the secondary. 

In the bottom row of Figure~\ref{orb}, we show a quantitative
comparison of disk gas orbits as a function of $a$, at $t=\,$3\,orbits,
corresponding to the~10, 15~and 20\,AU cases from left to right,
respectively. The maps consistently show that $e$ is a function of
both, $r_d$ and $a$. For a particular disk, material located at
small disk radii shows moderately higher $e$ than material located
at larger radii. Comparing disks with different $a$ we see that (i)
the magnitude of the average $e$ increases with increasing $a$ and (ii)
the angle between the semi-major axis of the orbits and the wind
velocity field decreases with $a$.  The latter is particularly clear
in orbits with small disk radii. This behaviour is in good agreement with our
findings in section~\ref{struct}; the angle is close to zero in the
20\,AU model (right-most panel).  

Finally, the color bars in the bottom panels of Figure~\ref{orb}
show disk gas position along the direction perpendicular to the
orbital plane; receding material is red whereas approaching
material is blue.  Note that we have chosen different color
limits in each panel in order to stress disk orbital features. The
color of the lines in the 10~and 15\,AU cases (left and middle
panels) consistently shows that material oscillates about the orbital
plane mildly with amplitudes no larger than 0.2~disk radii and frequencies
of order half~an orbit. The oscillation are much weaker in the
20\,AU case.  We do not see any correlation between the color of
the lines --i.e. the position about the orbital plane-- and the
wind velocity field direction (which points from the top left corner
to the bottom right one; Figure~\ref{orb} bottom row).

%%%%%%%%%%%%%%%%%%%%%%%%%%%%%%%%%%%%%%%%%%%%%%%%%%%%%%%%%%%%
\section{Discussion} \label{discu}

\subsection{Accretion onto the companion and implications for PPN/PN
phenomenology }
\label{acc}

\begin{figure}
\centering
  \includegraphics[width=\columnwidth]{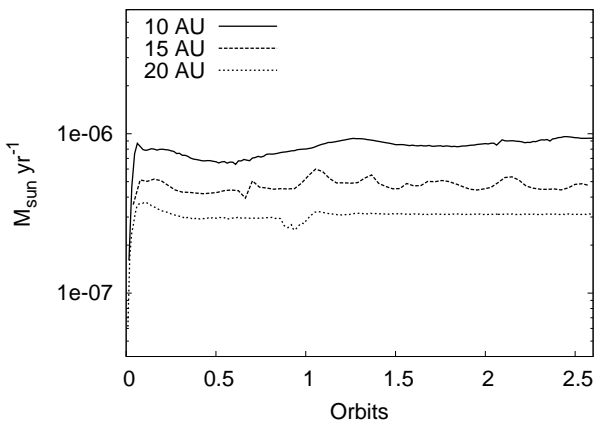}
\caption{Secondary accretion rate evolution. 
}
  \label{mdot}
\end{figure}

In Figure~\ref{mdot}, we show the evolution of the secondary's
accretion rate. The plots show the total gas flow into a region
within a 4\,cell radius from the star's center (section~\ref{sink})
as a function of time.  The plots demonstrate that after a steep
initial increase which lasts for about 0.05\,orbits the accretion
rate reaches a quasi-steady value with oscillations of a period
that decreases with $a$. These accretion histories are consistent
with those of \citet[][see their Figure~12]{borro}.  We note that
we see a small dip in the 20\,AU profile, close to 0.9\,orbits,
which corresponds to the disk formation time.

The analytic Bondi-Hoyle accretion rate
for a high velocity wind and negligible pressure conditions 
is given by
\begin{equation}
   \dot{m}_{BH} = \pi R_b^2 \rho({\bf x}_s) \sqrt{v_w^2+v_s^2},  
   \label{bhacc}
\end{equation}
\noindent where $R_b=\,$2$Gm_s\,(v_w^2+v_s^2)^{-1}$ is the modified
BHL radius (Edgar, 2004). For our~10, 15~and 20\,AU models,
$\dot{m}_{BH}$ takes the values of 
3.3, 1.8~and 1.2$\times\,$10$^{-6}\,$M$_{\odot}\,$yr$^{-1}$, respectively. Thus
the secondary accretion rates we find 
   in Figure~\ref{mdot}
	(0.85, 0.5~and 0.3$\times\,$10$^{-6}\,$M$_{\odot}\,$yr$^{-1}$,
	for $a=\,$10, 15 and 20\,AU, respectively) are
   $\dot{m}_{10AU} \sim\,$0.26$ \dot{m}_{BH,10}$,
   $\dot{m}_{15AU} \sim\,$0.28$ \dot{m}_{BH,15}$ and
   $\dot{m}_{20AU} \sim\,$0.25$ \dot{m}_{BH,20}$,
where the subscripts: $BH,10; BH,15; BH,20$ indicate the Bondi 
accretion rates at $10,15$ and $20$\,AU.

Our results appear to be consistent with those found by other authors.
For example in the models of \citet{borro}, the one with $a=\,$70\,AU
produced $\dot{M}_{acc}/\dot{M}_{wind} \sim$0.06 after 2 orbits.
While we have not run any simulations with such a large orbital separation,
these values are in line with the trend we observe in our results.
The semi-analytic models of
\citet{perets} yield $\dot{m}$ values that are also consistent with
ours. For example their Model 2 with $a=10$\,AU, $m_p=\,$3.0\,M$_{\odot}$,
$m_s=\,$1.0\,M$_{\odot}$ and ${\dot{M}}_p=\,$10$^{-5}$ M$_{\odot}$\,y$^{-1}$
yielded $\dot{M}_{acc} \sim\,$3$\times$10$^{-7}$.  We note that the
models of \citet{perets} are semi-analytic, therefore a direct comparison
between our work and theirs is not straightforward.

	The formation of these disks depends on 
	the specific angular momentum of the accreted mass, $j_a$
	(Soker \& Rappaport, 2000). It is important to consider 
	$j_a$ relative to the analytic net specific angular momentum of wind
	material that enters $r_b$, $j_{BH}$, which is given by 
   \begin{equation}
	   j_{BH} = \frac{1}{2} (\frac{2 \pi}{P} ) r_b^2,
	\label{angMom1}
	\end{equation}
   \noindent where $P$ is the orbital period.
   Our models show that %for $a=\,$10, 15 and~20\,AU, 
	$j_{a}/j_{BH}(a=\,$10)$=\,$0.35/2.52$=\,$0.13, 
	$j_{a}/j_{BH}(a=\,$15)$=\,$0.35/1.89$=\,$0.18 and 
	$j_{a}/j_{BH}(a=\,$20)$=\,$0.30/1.47$=\,$0.20. Here, we computed the numerators 
   using the accreted gas and calculating 
	a representative average value of the $z$ component (perpendicular
	to the orbital plane) of the cross product given by the radial
	distance from the secondary's center times the gas velocity. 
%j_bh (10,15,20AU) [cu]: 2.522874776, 1.891780013, 1.473590286, excel
   These figures are slightly higher, yet sensibly close, to analytical
	estimates of
	the specific angular momentum of isothermal high Mach number flows, for which
   $j_a/j_{BH} \simeq \,$0.1 
   (Livio et al., 1986; Ruffert, 1999). 

The results shown in Figure~\ref{mdot} are relevant for constraining
the class of models that can power PN outflows and, in particular,
the PPN outflows in the reflection nebulae thought to precede the
PN phase. The observations of Bujarrabal et al.~(2001) imply that
statistically, many PPN seem to harbor very powerful outflows whose
energies and estimated acceleration times give mechanical
luminosities as high as  4 $\times$ 10$^{36}$\,erg\,s$^{-1}$
(Blackman 2009, see also Huggins 2012). Can the accretion rates 
we find power such high luminosities?

If these high powered PPN outflows  do indeed result from accretion
disk around a secondary, then one typically expects no more than
10\% of the accretion rate to turn into mass outflow and the outflow
speed to be $\Omega r_A=Q v_K$ (e.g. Frank \& Blackman 2004), where $\Omega$
is the orbital speed at the inner edge of the disk, $r_A$ is the radius
of the Alfv\'en surface,  $v_K$ is the Keplerian speed at that
radius and $Q$ is a numerical factor determining the jet speed
as a multiple of the Keplerian speed (if the jet propagated into
a vacuum).  Although PPN jets would propagate into the stellar
envelope such that the observed jet speed would be reduced using
momentum conservation (Blackman 2009),  the mechanical luminosity
of the jet should be comparable to its naked value if energy is
conserved in the outflow. Accretion would provide a jet mechanical
luminosity of 
\begin{equation}
\begin{array}{l l}
   L_j & \sim {1\over 2}{{\dot M}_{ac}\over 10}Q^2 v_K^2 \\
	~   &=4\times 10^{36}\left({M_{ac}\over 5.6  \times
   10^{-5}{M_\odot/{\rm yr}}}\right)\left({Q\over 3}\right)^2 
   \left({v_K\over 440{km/s}}\right)^2,
   \label{lumin}
\end{array} 
\end{equation}
\noindent where we have
scaled $v_K$ to that of an orbit at the surface of a main sequence
star of 1\,$M_\odot$ and 1\,$R_\odot$ and taken an optimistic $Q=\,$3.
The accretion rates we find (see Figure~\ref{mdot}) would then  fall short by
$\sim \,$2 orders of magnitude if the secondary is a main sequence
star.  However, for a white dwarf, $v_K^2$ is $\sim \,$70 times larger
and the required accretion rate to produce the same mechanical
luminosity is reduced by the same factor, and becomes 8 $\times$
10$^{-7}$\,M$_\odot$\,yr$^{-1}$. This 
rate is not that dissimilar from the
accretion rate we found for the 10\,AU separation case. Note that
the estimates above depend on the value of $Q$ which is somewhat
uncertain and on the actual inner most radius of the disk used to
compute $v_K$. It could be that the disk does not extend all the way
to the star in which case $v_K$ in the above formula would be
reduced.

The fact that only a white dwarf companion could supply the outflow
powers for the highest powered PPN, highlights the need to  determine
the mechanical luminosity distribution function of PPN. If the
highest outflow powers are much more common than the expected
frequency of white dwarf companions within 10\,AU then the required
mechanism would likely require closer binary interactions and
perhaps even common envelope evolution. More work is needed to
assess the various modes of  accretion that can operate within a
common envelope, either onto the primary, or onto the secondary
\citep{2006MNRAS.370.2004N, 2011PNAS..108.3135N}.  It should be
noted that the late stage PN are less demanding with respect to
required accretion rates than the early PPN objects and it may be
that not all PPN are as demanding as the Bujarrabal et al.~(2001)
sample.

One example of a lower power object of interest  is the Red
Rectangle PPN.  In fact this object, reported in Bujarrabal et
al.~(2001), is one of the lower powered PPN presented therein.
Witt et al.~(2009) showed that the Red Rectangle is consistent
with accretion onto a main sequence companion  at binary separation
$<$1\,AU. The jet has a mechanical luminosity  $\sim \,$ 7$\times$
10$^{33}$\,erg\,s$^{-1}$.  Witt et al. (2009) plausibly
argue from presumed evidence
for disk emission that the disk around the companion is accreting
via Roche lobe overflow from the primary at a rate 2--5$\times$
10$^{-5}$\,M$_\odot$\,yr$^{-1}$.
However, these authors argue that  the
jet outflow speed would be launched at 100\,km\,s$^{-1}$ given
this accretion rate, which is significantly below the escape
speed at the inner edge of the disk if the disk were to extend
all the way to the star.   Such an  accretion rate could certainly
power the observed jets in the Red Rectangle.

In the future, a larger parameter survey covering a systematic range
of  grid resolution setups and accreting radii about the secondary 
(sink particle) --gravity softening radii-- would be helpful both, to
further constrain the connection between core accretion and jet
launch in binary-formed disks and to better characterize numerical
artifacts.

\section{CONCLUSIONS AND SUMMARY}
\label{conclu}

We have performed the highest resolution simulations to date of
wind accretion and disk formation in an evolved star binary system.
We simulated the wind accretion process at three different orbital
radii, 10, 15 and 20\,AU, and found accretion disk formation in all
cases. We have also derived an 
   accretion line
impact parameter criterion that
determines the minimum required resolution for such simulations to
produce physically consistent results. 

In our derivation we assume the wind experiences a
balance of radiation pressure and gravitational forces
from the primary. This is equivalent to assuming the
secondary orbits beyond the wind's acceleration zone.
Operationally this means a differential acceleration between
the wind and the secondary occurs since the secondary does
feel the full gravity from the primary. It is this
differential acceleration which creates the accretion line
impact parameter.  We note however that even in the wind
acceleration zone the radiation pressure force on the gas
(but not on the secondary) would produce such a differential
gravitational acceleration.

The accretion line impact parameter $b$ measures
the perpendicular (retarded) distance from the
stagnation point in the BH accretion stream to the accretor and as
such is a measure of the distance at which material from the wind
is captured into orbit around the secondary.  
We have compared the validity of this parameter to
other formulations of a ``disk formation'' radius and found that in the regime of applicability $b$ is the larger
of the expressions and hence represents the outer most distance at which a disk will form.

Additionally, accretion disks formed in our simulations at~10 and~15\,AU
 incur mass oscillations which appear to result
from ram pressure stripping of the outer disk radii. The 15\,AU
oscillations become weaker as the wind ram pressure decreases with
orbital radius 
(due to the wind density's $r^{-2}$ density profile (\ref{densw}). 
The 20\,AU disks, showed no oscillations, consistent
with an even weaker wind ram pressure.

The disk accretion rates evolve to be quasi-steady after 1$/$4\,orbits.
The accretion rates are lower for larger orbital radius, taking
values~8, 5~to 3~$\times$10$^{-7}\,$M$_{\odot}\,$yr$^{-1}$,
for the~10, 15~and 20\,AU cases, respectively.  We note that
for the secondary's mass that we explore here, the simplest version
of Bondi-Hoyle accretion model (steady uniform parallel wind flow)
only applies truly at large separations; only in such cases
is the angle between the accretion flow and the wind zero. We see
this angle becoming $<\,$5$^{\circ}$ for the 20\,AU case, but it
is non-negligible for the smaller radii runs.

While the basic problem of accretion from a wind studied
herein has broad applicability in a number of contexts, we chose
the parameters to be consistent with those of a stellar mass companion
accreting from an AGB wind. 
This is motivated by the need to
identify constraints on the mode of accretion that might be
needed to explain the accretion powered jets and outflows of PPN and PN.  
Toward this end, we find that the accretion rates from our simulations
fall  short of what is needed to be able to power the most powerful
young bipolar PPN unless the companion is a white dwarf.  Closer
binary interactions, which includes  common envelope evolution, are
likely  an important player for some PPN objects but; our present
work focused only on binary radii at 10\,AU and larger separation in
order to be be largely outside of the radiative wind acceleration
zone. It is useful to compare our work with other studies of wind capture
accretion in binary systems.  Theuns \& Jorissen~(1993) used SPH
simulations and despite low resolution by today's standards were
able to capture the basic features of a bow-shock and disk formation.
In the seminal work by Mastrodemos \& Morris~(1999) SPH methods
were again used to systematically explore the formation of accretion
disks.  Their work implied that disks could form out to distances
of at least 21\,AU.  Their 3-D models also hinted at the formation
of significant warps in the disks which would be important as these
might drive precession in any jets which form from the disks.  The
work of \citet{borro} found disks forming out to radii of 70\,AU.
If such a result where to hold it would be important for establishing
the population of binaries for which disk winds might power jets.
Unfortunately the \citet{borro} study was carried out in 2-D only
so it not possible with our work to confirm their outer limits on
disk formation.  Of equal importance was the role of the inner AGB
wind launching radius $R_w$  in their studies.  \citet{borro} allowed
the inner wind launching radius to be a free parameter.  The
justification for this was the possibility of large wind acceleration
zones for AGB stars.  In their simulations \citet{borro} found that
as $R_w$ increased the flow pattern became more complex appearing
as a mix between Roche Lobe overflow and wind capture.  Such a mix
of accretion modes was also implied by \citet{mohamed07}. In
\citet{perets} semi-analytic methods also found disks forming at
large binary separations ($a = \,$100\,AU).  If these results can be
verified in full 3-D fluid dynamical simulations they would also
enlarge the population range of outflow bearing systems (though
these outflows would likely be of low power as explained in the
previous section). Taken as a whole, understanding these possibilities
in greater detail will be essential for understanding the evolutionary
pathways and outflow generation of PPN and PN systems.

Finally we note that our study provides new insights into the
Bondi-Hoyle accretion in binary systems.  We find for example that
the stagnation region in which the accretion ``line'' separates into
flows directed towards and away from the accretor is best viewed
as a vortex tube as instreaming material flows along helical paths
towards the secondary star.  In addition we find the flows (and
most important the disks) which form via binary BH flows appear
quite stable \citep{blondin09,blondin12}. Only small
amplitude fluctuations in the orientation of different disk annuli
are seen and these would appear to be too small to drive significant
jet precession.
In conclusion, more work is needed to explore the consequences of interacting
binaries for a wider range of orbital radii. And more observations
are needed to determine the mechanical luminosity function of PPN
jets to assess what fraction have the high powered jets that are
typical of the Bujarrabal et al. (2001) PPN sample.

\section*{Acknowledgments}
Financial support for this project was provided by the Space Telescope
Science Institute grants HST-AR-11251.01-A and HST-AR-12128.01-A;
by the National Science Foundation under award AST-0807363; by the
Department of Energy under award DE-SC0001063. JN is supported by an NSF Astronomy
and Astrophysics Postdoctoral Fellowship under award AST-1102738
and by NASA HST grant AR-12146.01-A. This work used the Extreme
Science and Engineering Discovery Environment (XSEDE), which is
supported by National Science Foundation grant number OCI-1053575.

\bsp

\label{lastpage}
 
\end{document}